\pdfoutput=1
\documentclass[12pt,a4paper]{article}

\usepackage{ifthen}
\newboolean{pdflatex}
\setboolean{pdflatex}{true}

\newboolean{articletitles}
\setboolean{articletitles}{true}

\newboolean{uprightparticles}
\setboolean{uprightparticles}{false}

\def\paperauthors{LHCb collaboration}
\def\paperasciititle{A measurement of $\Delta \Gamma_s$}
\def\papertitle{A measurement of $\Delta \Gamma_s$}
\def\paperkeywords{{High Energy Physics}, {LHCb}}
\def\papercopyright{\the\year\ CERN for the benefit of the LHCb collaboration}
\def\paperlicence{CC BY 4.0 licence}
\def\paperlicenceurl{https://creativecommons.org/licenses/by/4.0/}

\usepackage[top=1in, bottom=1.25in, left=1in, right=1in]{geometry}

\usepackage{microtype}
\usepackage{lineno}
\usepackage{xspace}
\usepackage{caption}

\usepackage{graphicx}
\usepackage{color}
\usepackage{colortbl}

\usepackage{amsmath}
\usepackage{amssymb}
\usepackage{amsfonts}
\usepackage{upgreek}

\newcommand*\patchAmsMathEnvironmentForLineno[1]{
\expandafter\let\csname old#1\expandafter\endcsname\csname #1\endcsname
\expandafter\let\csname oldend#1\expandafter\endcsname\csname
end#1\endcsname
 \renewenvironment{#1}
   {\linenomath\csname old#1\endcsname}
   {\csname oldend#1\endcsname\endlinenomath}
}
\newcommand*\patchBothAmsMathEnvironmentsForLineno[1]{
  \patchAmsMathEnvironmentForLineno{#1}
  \patchAmsMathEnvironmentForLineno{#1*}
}
\AtBeginDocument{
\patchBothAmsMathEnvironmentsForLineno{equation}
\patchBothAmsMathEnvironmentsForLineno{align}
\patchBothAmsMathEnvironmentsForLineno{flalign}
\patchBothAmsMathEnvironmentsForLineno{alignat}
\patchBothAmsMathEnvironmentsForLineno{gather}
\patchBothAmsMathEnvironmentsForLineno{multline}
\patchBothAmsMathEnvironmentsForLineno{eqnarray}
}

\usepackage{hyperxmp}

\usepackage[pdftex,
            pdfauthor={\paperauthors},
            pdftitle={\paperasciititle},
            pdfkeywords={\paperkeywords},
            pdfcopyright={Copyright (C) \papercopyright},
            pdflicenseurl={\paperlicenceurl}]{hyperref}

\usepackage[colorinlistoftodos,textsize=scriptsize]{todonotes}

\usepackage[bottom,flushmargin,hang,multiple]{footmisc}

\usepackage[all]{hypcap}

\usepackage{xspace}
\usepackage{upgreek}

\def\lhcb   {\mbox{LHCb}\xspace}

\def\MagUp {\mbox{\em Mag\kern -0.05em Up}\xspace}

\ifthenelse{\boolean{uprightparticles}}
{

 \def\Peta        {\ensuremath{\upeta}\xspace}

 \def\Pmu         {\ensuremath{\upmu}\xspace}

 \def\Ppi         {\ensuremath{\uppi}\xspace}

 \def\Pphi        {\ensuremath{\upphi}\xspace}

 \def\Ppsi        {\ensuremath{\uppsi}\xspace}

 \def\PDelta      {\ensuremath{\Delta}\xspace}
 \def\PXi         {\ensuremath{\Xi}\xspace}
 \def\PLambda     {\ensuremath{\Lambda}\xspace}
 \def\PSigma      {\ensuremath{\Sigma}\xspace}
 \def\POmega      {\ensuremath{\Omega}\xspace}
 \def\PUpsilon    {\ensuremath{\Upsilon}\xspace}
 \let\oldPi\Pi
 \def\PPi         {\ensuremath{\oldPi}\xspace}

 \def\PB      {\ensuremath{\mathrm{B}}\xspace}
 
 \def\PD      {\ensuremath{\mathrm{D}}\xspace}

 \def\PJ      {\ensuremath{\mathrm{J}}\xspace}
 \def\PK      {\ensuremath{\mathrm{K}}\xspace}

 \def\Pb      {\ensuremath{\mathrm{b}}\xspace}
 \def\Pc      {\ensuremath{\mathrm{c}}\xspace}

 \def\Pi      {\ensuremath{\mathrm{i}}\xspace}

 \def\Ps      {\ensuremath{\mathrm{s}}\xspace}

 \def\thebaroffset{0.0em}
}
{

 \def\Peta        {\ensuremath{\eta}\xspace}

 \def\Pmu         {\ensuremath{\mu}\xspace}

 \def\Ppi         {\ensuremath{\pi}\xspace}

 \def\Pphi        {\ensuremath{\phi}\xspace}

 \def\Ppsi        {\ensuremath{\psi}\xspace}
 
 \mathchardef\PDelta="7101
 \mathchardef\PXi="7104
 \mathchardef\PLambda="7103
 \mathchardef\PSigma="7106
 \mathchardef\POmega="710A
 \mathchardef\PUpsilon="7107
 \mathchardef\PPi="7105
 
 \def\PB      {\ensuremath{B}\xspace}
 
 \def\PD      {\ensuremath{D}\xspace}

 \def\PJ      {\ensuremath{J}\xspace}
 \def\PK      {\ensuremath{K}\xspace}

 \def\Pb      {\ensuremath{b}\xspace}
 \def\Pc      {\ensuremath{c}\xspace}

 \def\Pi      {\ensuremath{i}\xspace}

 \def\Ps      {\ensuremath{s}\xspace}

 \def\thebaroffset{0.18em}
}
\newcommand{\offsetoverline}[2][\thebaroffset]{\kern #1\overline{\kern -#1 #2}}

\makeatletter
\ifcase \@ptsize \relax
  \newcommand{\miniscule}{\@setfontsize\miniscule{4}{5}}
\or
  \newcommand{\miniscule}{\@setfontsize\miniscule{5}{6}}
\or
  \newcommand{\miniscule}{\@setfontsize\miniscule{5}{6}}
\fi
\makeatother

\DeclareRobustCommand{\optbar}[1]{\shortstack{{\miniscule (\rule[.5ex]{1.25em}{.18mm})}
  \\ [-.7ex] $#1$}}

\def\mup        {{\ensuremath{\Pmu^+}}\xspace}
\def\mun        {{\ensuremath{\Pmu^-}}\xspace}

\def\mumu       {{\ensuremath{\Pmu^+\Pmu^-}}\xspace}

\def\squark    {{\ensuremath{\Ps}}\xspace}

\def\cquark    {{\ensuremath{\Pc}}\xspace}

\def\bquark    {{\ensuremath{\Pb}}\xspace}

\def\pion   {{\ensuremath{\Ppi}}\xspace}
\def\piz    {{\ensuremath{\pion^0}}\xspace}
\def\pip    {{\ensuremath{\pion^+}}\xspace}
\def\pim    {{\ensuremath{\pion^-}}\xspace}

\def\kaon    {{\ensuremath{\PK}}\xspace}

\def\KorKbar {\kern \thebaroffset\optbar{\kern -\thebaroffset \PK}{}\xspace}

\def\Kp      {{\ensuremath{\kaon^+}}\xspace}
\def\Km      {{\ensuremath{\kaon^-}}\xspace}

\newcommand{\etaz}{\ensuremath{\Peta}\xspace}
\newcommand{\etapr}{\ensuremath{\Peta^{\prime}}\xspace}
\newcommand{\phiz}{\ensuremath{\Pphi}\xspace}

\def\D       {{\ensuremath{\PD}}\xspace}

\def\DorDbar {\kern \thebaroffset\optbar{\kern -\thebaroffset \PD}\xspace}

\def\Dp      {{\ensuremath{\D^+}}\xspace}
\def\Dm      {{\ensuremath{\D^-}}\xspace}

\def\DpDm    {\ensuremath{\Dp {\kern -0.16em \Dm}}\xspace}

\def\B       {{\ensuremath{\PB}}\xspace}
\def\Bbar    {{\ensuremath{\offsetoverline{\PB}}}\xspace}

\def\BorBbar {\kern \thebaroffset\optbar{\kern -\thebaroffset \PB}\xspace}
\def\Bz      {{\ensuremath{\B^0}}\xspace}

\def\Bd      {{\ensuremath{\B^0}}\xspace}

\def\BdorBdbar {\kern \thebaroffset\optbar{\kern -\thebaroffset \Bd}\xspace}
\def\Bu      {{\ensuremath{\B^+}}\xspace}

\def\Bp      {{\ensuremath{\Bu}}\xspace}

\def\Bs      {{\ensuremath{\B^0_\squark}}\xspace}
\def\Bsb     {{\ensuremath{\Bbar{}^0_\squark}}\xspace}
\def\BsorBsbar {\kern \thebaroffset\optbar{\kern -\thebaroffset \Bs}\xspace}

\def\jpsi     {{\ensuremath{{\PJ\mskip -3mu/\mskip -2mu\Ppsi}}}\xspace}
\def\psitwos  {{\ensuremath{\Ppsi{(2S)}}}\xspace}

\def\Y#1S{\ensuremath{\PUpsilon{(#1S)}}\xspace}

\def\Lz          {{\ensuremath{\PLambda}}\xspace}

\def\LorLbar     {\kern \thebaroffset\optbar{\kern -\thebaroffset \PLambda}\xspace}

\def\Lb           {{\ensuremath{\Lz^0_\bquark}}\xspace}

\newcommand{\decay}[2]{\ensuremath{#1\!\to #2}\xspace}

\def\to                 {\ensuremath{\rightarrow}\xspace}

\def\CP                {{\ensuremath{C\!P}}\xspace}

\newcommand{\phis}{{\ensuremath{\phi_{\squark}}}\xspace}

\def\AT#1     {\ensuremath{A_{\mathrm{T}}^{#1}}\xspace}

\def\C#1      {\ensuremath{\mathcal{C}_{#1}}\xspace}
\def\Cp#1     {\ensuremath{\mathcal{C}_{#1}^{'}}\xspace}
\def\Ceff#1   {\ensuremath{\mathcal{C}_{#1}^{\mathrm{(eff)}}}\xspace}
\def\Cpeff#1  {\ensuremath{\mathcal{C}_{#1}^{'\mathrm{(eff)}}}\xspace}
\def\Ope#1    {\ensuremath{\mathcal{O}_{#1}}\xspace}
\def\Opep#1   {\ensuremath{\mathcal{O}_{#1}^{'}}\xspace}

\newcommand{\aunit}[1]{\ensuremath{\text{\,#1}}}

\newcommand{\tev}{\aunit{Te\kern -0.1em V}\xspace}
\newcommand{\gev}{\aunit{Ge\kern -0.1em V}\xspace}
\newcommand{\mev}{\aunit{Me\kern -0.1em V}\xspace}
\newcommand{\kev}{\aunit{ke\kern -0.1em V}\xspace}
\newcommand{\ev}{\aunit{e\kern -0.1em V}\xspace}

\newcommand{\mevc}{\ensuremath{\aunit{Me\kern -0.1em V\!/}c}\xspace}
\newcommand{\gevc}{\ensuremath{\aunit{Ge\kern -0.1em V\!/}c}\xspace}
\newcommand{\mevcc}{\ensuremath{\aunit{Me\kern -0.1em V\!/}c^2}\xspace}
\newcommand{\gevcc}{\ensuremath{\aunit{Ge\kern -0.1em V\!/}c^2}\xspace}

\def\fb   {\ensuremath{\aunit{fb}}\xspace}
\def\invfb   {\ensuremath{\fb^{-1}}\xspace}

\def\ns   {\ensuremath{\aunit{ns}}\xspace}
\def\ps   {\ensuremath{\aunit{ps}}\xspace}
\def\fs   {\aunit{fs}}

\def\invns{\ensuremath{\ns^{-1}}\xspace}

\newcommand{\chisq}{\ensuremath{\chi^2}\xspace}

\newcommand{\chisqip}{\ensuremath{\chi^2_{\text{IP}}}\xspace}

\def\gsim{{~\raise.15em\hbox{$>$}\kern-.85em
          \lower.35em\hbox{$\sim$}~}\xspace}
\def\lsim{{~\raise.15em\hbox{$<$}\kern-.85em
          \lower.35em\hbox{$\sim$}~}\xspace}

\def\pt         {\ensuremath{p_{\mathrm{T}}}\xspace}

\def\ptot       {\ensuremath{p}\xspace}

\def\rad{\aunit{rad}\xspace}

\def\evtgen     {\mbox{\textsc{EvtGen}}\xspace}

\def\geant      {\mbox{\textsc{Geant4}}\xspace}

\def\photos     {\mbox{\textsc{Photos}}\xspace}

\def\pythia     {\mbox{\textsc{Pythia}}\xspace}

\def\tell1  {TELL1\xspace}
\def\ukl1   {UKL1\xspace}

\newcommand{\lhcborcid}[1]{\href{https://orcid.org/#1}{\hspace*{0.1em}\raisebox{-0.45ex}{\includegraphics[width=1em]{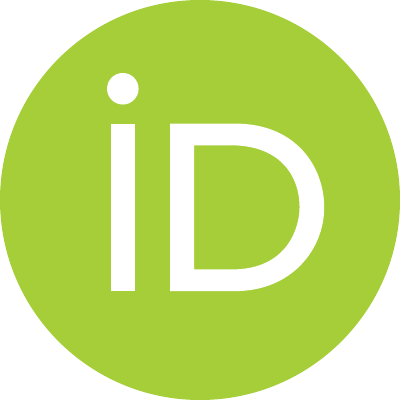}}}}

\usepackage{cite}
\usepackage{mciteplus}

\begin{document}

\renewcommand{\thefootnote}{\fnsymbol{footnote}}
\setcounter{footnote}{1}

\begin{titlepage}
\pagenumbering{roman}

\vspace*{-1.5cm}
\centerline{\large EUROPEAN ORGANIZATION FOR NUCLEAR RESEARCH (CERN)}
\vspace*{1.5cm}
\noindent
\begin{tabular*}{\linewidth}{lc@{\extracolsep{\fill}}r@{\extracolsep{0pt}}}
\vspace*{-1.5cm}\mbox{\!\!\!\includegraphics[width=.14\textwidth]{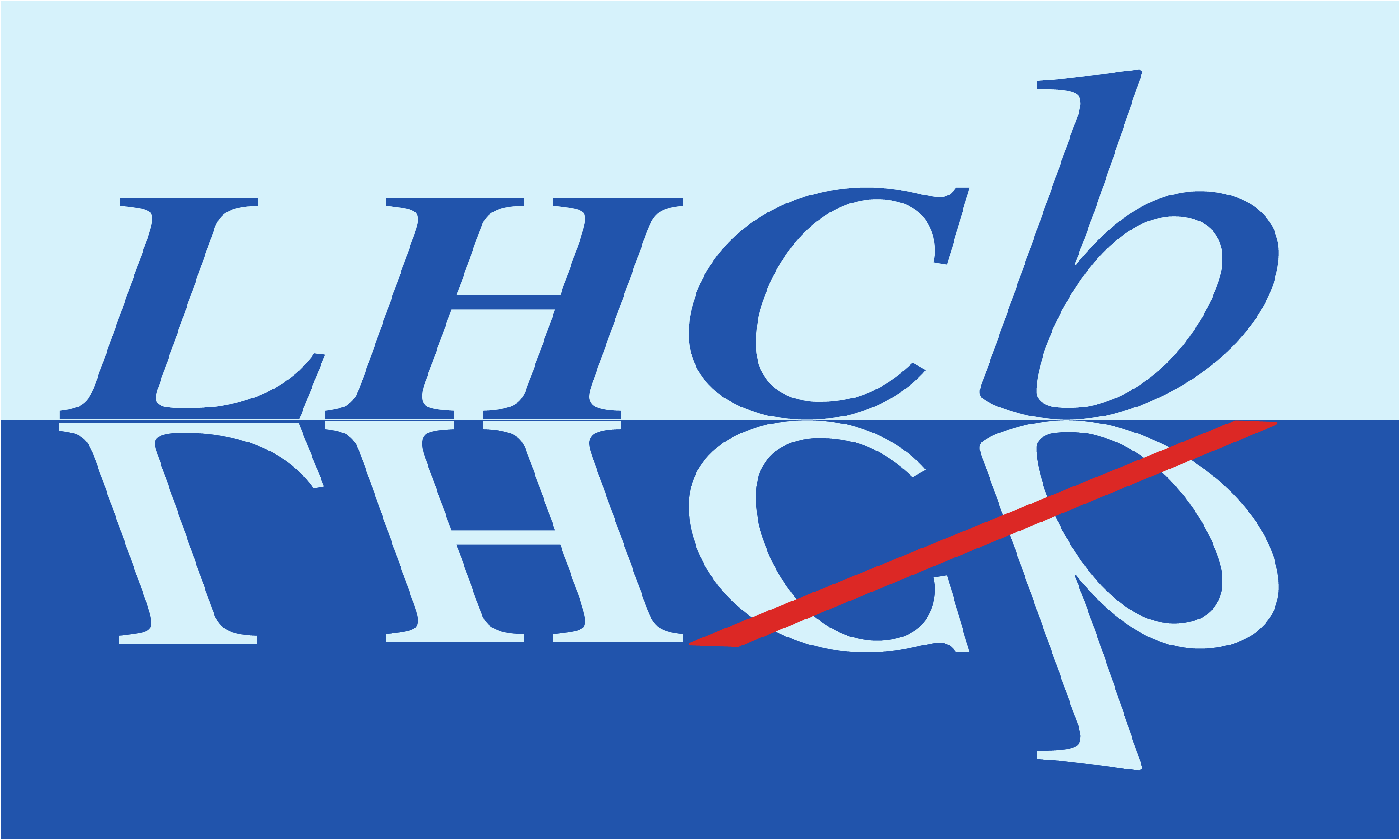}} & &
\\
 & & CERN-EP-2023-218 \\
 & & LHCb-PAPER-2023-025 \\
 & & \today \\

\end{tabular*}

\vspace*{4.0cm}

{\normalfont\bfseries\boldmath\huge
\begin{center}
  \papertitle
\end{center}
}

\vspace*{2.0cm}

\begin{center}
\paperauthors\footnote{Authors are listed at the end of this paper.}
\end{center}

\vspace{\fill}

\begin{abstract}
  \noindent
  Using a dataset corresponding to $9 \invfb$ of integrated luminosity collected with the LHCb detector between 2011 and 2018 in proton-proton collisions, the decay-time distributions of the decay modes $B_s^0 \rightarrow \jpsi \eta'$ and $B_s^0 \rightarrow \jpsi \pip \pim$ are studied. The
  decay-width difference between the light and heavy mass eigenstates of the $\Bs$ meson is measured to be $\Delta \Gamma_s = 0.087 \pm 0.012 \pm 0.009 \, \mathrm{ps}^{-1}$, where the first uncertainty is statistical and the second systematic.

\end{abstract}

\vspace*{2.0cm}

\begin{center}
  Published in JHEP 05 (2024) 253
\end{center}

\vspace{\fill}

{\footnotesize
\centerline{\copyright~\papercopyright. \href{\paperlicenceurl}{\paperlicence}.}}
\vspace*{2mm}

\end{titlepage}

\newpage
\setcounter{page}{2}
\mbox{~}

\renewcommand{\thefootnote}{\arabic{footnote}}
\setcounter{footnote}{0}

\cleardoublepage

\pagestyle{plain}
\setcounter{page}{1}
\pagenumbering{arabic}

\section{Introduction}
\label{sec:Introduction}
Measurements of $\Bs-\Bsb$ mixing parameters are powerful tests of the Standard Model (SM) of particle physics. The $\Bs$ and $\Bsb$ flavour eigenstates can be expressed as linear combinations of the heavy (H) and light (L) mass eigenstates. The decay-width difference, $\Delta \Gamma_{s} = \Gamma_{\rm L} - \Gamma_{\rm H}$, is predicted to be in the range $(7.6 - 9.2) \times 10^{-2} \ps^{-1}$ depending on the choice of the renormalisation scheme\cite{Asatrian:2020zxa,Davies:2019gnp,Lenz:2019lvd,Gerlach:2022wgb}. Experimentally, the golden channel for the measurement of the $\Bs-\Bsb$ mixing parameters is the decay $\Bs \to \jpsi \phi$ since it gives a clean experimental signature and is relatively abundant. Measurements of $\Delta \Gamma_s$ using the $\Bs \to \jpsi \phi$ decay have been reported by the ATLAS \cite{ATLAS:2020lbz}, CMS \cite{CMS:2020efq} and LHCb \cite{LHCb-PAPER-2023-016} collaborations. Though precise, these measurements are in tension with one another, which results in a large scale factor on the $\Delta\Gamma_s$ uncertainty in the HFLAV average \cite{HFLAV21}. This has motivated independent measurements in other decay modes such as $\Bs \to J/\psi K^+ K^-$ above the $\phi$ mass-region ~\cite{LHCB-PAPER-2017-008}, and $\Bs \to \psi(2S) \phi$~\cite{LHCB-PAPER-2016-027} .

The formalism used in this analysis is similar to that proposed in Ref.~\cite{Fleischer:2011cw} to measure the $\Bs$ mixing phase, $\phis$.
 The small value of $\phi_s$ measured experimentally means that \CP-even modes determine the light mass eigenstate lifetime ($\tau_{\rm L} = 1/\Gamma_{\rm L}$) while \CP-odd modes measure the heavy mass eigenstate lifetime ($\tau_{\rm H} = 1/\Gamma_{\rm H}$), within 0.2$\fs^{-1}$~\cite{Fleischer:2011cw}. Thus, $\Delta \Gamma_{s}$ can be determined from the decay-width difference between a \CP-odd and a \CP-even $\Bs$ mode.

In this paper, $\Delta \Gamma_{s}$ is measured from the decay-width difference between the \CP-even decay of $\Bs \to \jpsi \etapr$, with subsequent decays $\jpsi \to \mu^+ \mu^-$ and $\etapr \to \rho^0 \gamma$, and the  $\Bs \to \jpsi \pip \pim$ decay mode with the subsequent decay $J/\psi \to \mu^+ \mu^-$. For the latter mode, the dipion mass is required to be within $90 \mevcc$  of the known $f_0(980)$ mass \cite{PDG2022} to select predominantly \CP-odd decays \cite{LHCb-PAPER-2013-069}. The study is performed using the full dataset collected by the LHCb collaboration in proton-proton ($pp$) collisions between 2011 and 2018. This corresponds to an integrated luminosity of 9 \invfb collected at centre-of-mass energies of $\sqrt{s}$ = 7, 8 and 13 TeV.

The analysis adopts a similar strategy to that used to place a limit on the corresponding parameter in the $\Bd$-system, $\Delta \Gamma_d$, by the ATLAS collaboration \cite{Aaboud:2016bro}. If \CP violation is negligible, the time-dependent decay rate to a $\CP$-eigenstate, $f$, is
\begin{equation}
\Gamma (\Bs (t) \rightarrow f) \propto e^{-\Gamma_s t} \left[ \textrm{cosh}\left(\frac{\Delta \Gamma_s t}{2}\right)  \, +  \, \eta_{\CP} \,
\textrm{sinh}\left(\frac{\Delta \Gamma_s t}{2}\right) \right] \, ,
\label{eq:even}
\end{equation}
where $\Gamma_s = (\Gamma_{\rm H} + \Gamma_{\rm L})/2$ and $\eta_{CP}$ is $-1$ for a \CP-even state and 1 for a \CP-odd state. Integrating Eq.~\ref{eq:even} over a time bin $[t_1,t_2]$ gives
\begin{equation}
N_{\rm L} \propto \left[ \frac{e^{t(\frac{-\Delta \Gamma_s}{2} -\Gamma_s)}}{\Gamma_s + \frac{\Delta \Gamma_s}{2}} \right]^{t_2}_{t_1} ,
\end{equation}
and
\begin{equation}
N_{\rm H} \propto \left[ \frac{e^{t(\frac{\Delta \Gamma_s}{2} -\Gamma_s)}}{\Gamma_s - \frac{\Delta \Gamma_s}{2}} \right]^{t_2}_{t_1},
\end{equation}
where $N_{\rm L}$ and $N_{\rm H}$ are the yields of the \CP-even and \CP-odd modes in that interval. The ratio of the yields in the interval is then
\begin{equation}
R_i = \frac{N_{\rm L}}{N_{\rm H}} \propto \frac{\left[e^{-\Gamma_s t(1 +y)} \right]^{t_2}_{t_1}}{ \left[e^{-\Gamma_s t(1-y)} \right]^{t_2}_{t_1} }\cdot \frac{(1-y)}{(1 + y)},
\label{eq:main}
\end{equation}
where $y = \Delta \Gamma_s/ 2\Gamma_s$. Experimentally, the determination of $R_i$ requires the observed yields $N^{\textrm{RAW}}_{\textrm{L,H}}$ to be corrected by the relative efficiency in each decay time bin, $A_i$. By writing
\begin{equation}
R_i = A_i \cdot \frac{N^{\textrm{RAW}}_{\rm L}}{N^{\textrm{RAW}}_{\rm H}},
\end{equation}
$\Delta \Gamma_{s}$ is determined from a $\chisq$ minimization of Eq.~\ref{eq:main} with $\Delta \Gamma_{s}$ and an arbitrary normalization factor as free parameters. Though Eq.~\ref{eq:main} depends on $\Gamma_s$, this dependence is weak and vanishes in the limit of narrow decay-time bins.

The number of bins and their ranges are chosen using simulation and pseudoexperiments so as to minimise the uncertainty on $\Delta \Gamma_{s}$. This results in eight bins (Table~\ref{tab:binning_scheme}) with similar yields expected in each. The lower decay-time limit of $0.5 \ps$ is chosen since above this value the time-acceptance is relatively flat. Above $10 \ps$ the expected yield for the $\Bs \to \jpsi \etapr$ decay mode is negligible. Since the last bins are relatively broad, the optimal value of $t$ used to evaluate $A_i$ within a bin needs to be considered \cite{Lafferty:1994cj}. Using pseudoexperiments, the barycentre of the bin calculated using an exponential decay-time model with $\Gamma = \Gamma_s$ is found to minimise the bias on $\Delta \Gamma_s$.
\begin{table}[htb!]
    \centering
      \caption{\small Decay-time binning scheme.}
    \label{tab:binning_scheme}
    \begin{tabular}{cc}
        Number  & Interval [ps]\\
        \hline
        1           & 0.5 -- 0.7\\
        2           & 0.7 -- 0.9\\
        3           & 0.9 -- 1.2\\
        4           & 1.2 -- 1.5\\
        5           & 1.5 -- 2.0\\
        6           & 2.0 -- 2.5\\
        7           & 2.5 -- 3.5\\
        8           & \phantom{0}3.5 - 10.0\\
    \end{tabular}
\end{table}

\section{Detector and simulation}
\label{sec:detector}
The \lhcb detector~\cite{LHCb-DP-2012-002,LHCb-DP-2014-002} is a single-arm forward spectrometer covering the \mbox{pseudorapidity} range $2<\eta <5$, designed for the study of particles containing \bquark or \cquark quarks. It includes a high-precision tracking system consisting of a silicon-strip vertex detector (VELO) surrounding the $pp$ interaction region, a large-area silicon-strip detector (TT) located upstream of a dipole magnet with a bending power of approximately $4{\mathrm{\,Tm}}$, and three stations of silicon-strip detectors and straw drift tubes placed downstream of the magnet. The tracking system provides a measurement of the momentum, \ptot, of charged particles with a relative uncertainty that varies from 0.5\,\% at low momentum to 1.0\,\% at $200 \gevc$. Large samples of  $\jpsi\to\mup\mun$ and $B^+ \to \jpsi K^+$ decays, collected concurrently with the dataset used in this analysis, are used to calibrate the momentum scale of the spectrometer~\cite{LHCb-PAPER-2013-011}. The relative uncertainty on the momentum scale is $3 \times 10^{-4}$. For $\bquark$-hadron decay modes such as $B^+ \to \jpsi K^+$, the mass resolution agrees between data and simulation to better than 10\,\%.

Various charged hadrons are distinguished using information from two ring-imaging Cherenkov (RICH) detectors. In addition, photons, electrons, and hadrons are identified by a calorimeter system consisting of scintillating-pad and preshower detectors, an electromagnetic and a hadronic calorimeter. The calorimeter response is calibrated
using samples of $\pi^0 \rightarrow \gamma \gamma$ decays \cite{LHCb-DP-2020-001}. Muons are identified by a system composed of alternating layers of iron and multiwire proportional chambers.

The online event selection is performed by a trigger, which consists of a hardware stage followed by a two-level software stage \cite{LHCb-DP-2019-001}. For the Run 2 dataset, the alignment, and calibration of the detector is performed in near real-time such that the results are used in the software trigger~\cite{Borghi:2017hfp}. The same alignment and calibration information is propagated to the offline reconstruction, ensuring consistent information between the trigger and offline software. The first stage of the software trigger performs a partial event reconstruction and requires events to have two well-identified oppositely charged muons with an invariant mass larger than $2.7 \gevcc$ without biasing the lifetime distribution. The second stage performs a
full event reconstruction.  Events are retained for further
processing if they contain a displaced $\jpsi \rightarrow \mu^+ \mu^-$
candidate. The  $\jpsi$  decay vertex is required to be well separated
from each reconstructed primary vertex (PV) of the $pp$ interaction by requiring the distance between the PV and the decay vertex  divided by its
uncertainty, the decay-length significance, to be greater than three. This introduces a nonuniform
efficiency for $\bquark$-hadron candidates that have a decay time less than
$\sim 0.4 \ps$.

Simulated $pp$ collisions are generated using
\pythia~\cite{Sjostrand:2007gs}  with a specific
\lhcb configuration~\cite{LHCb-PROC-2010-056}.  Decays of hadronic
particles are described by \evtgen~\cite{Lange:2001uf}, in which
final-state radiation is generated using
\photos~\cite{Golonka:2005pn}. The interaction of the generated
particles with the detector, and its response, are implemented using the \geant toolkit~\cite{Allison:2006ve,
  *Agostinelli:2002hh} as described in Ref.~\cite{LHCb-PROC-2011-006}. Several sources of background are also studied using the RapidSim fast simulation package \cite{Cowan:2016tnm}.

\section{Event selection}
\label{sec:selection}

The offline selection for both the $\Bs \rightarrow \jpsi \etapr$ and $\Bs \rightarrow \jpsi \pip \pim$ modes starts from a pair of oppositely charged particles, identified as muons, that form a common decay vertex. Each muon must have a transverse momentum ($\pt$) greater than $500 \mevc$ and good track quality. The invariant mass of the \mumu~pair is required to be
within $\pm50 \mevcc$ of the known \jpsi~mass \cite{PDG2022}.

Photon candidates are selected from well-identified neutral clusters reconstructed in the electromagnetic
calorimeter that have a transverse energy in excess of $500 \mev$. Candidate $\rho^0$ mesons are reconstructed from pairs of tracks with opposite charge that are identified as pions by the RICH detectors.  The invariant mass of the dipion pair is required to be within the range 620--930\mevcc. To select the $\Bs \rightarrow \jpsi \pip \pim$ decay,  pairs of pion candidates with opposite charge are combined in a similar manner. In this case, the invariant mass of the dipion pair is required to be within $90 \mevcc$ of the world average $f_{0}(980)$ mass \cite{PDG2022} and the scalar sum of the $\pt$  values for the two pions must be greater than $1 \gevc$.

Candidate $\etapr \to\rho^0 \gamma$ decays are formed by combining the selected  $\rho^0$ and $\gamma$ candidates. The invariant mass of the combination is required to be within $50 \mevcc$ of the known $\etapr$ mass \cite{PDG2022} and the $\pt$ of the candidate must be larger than $2 \gevc$.

Candidate $\Bs$ decays are formed from the selected $\jpsi$ and $\etapr$ candidates, or dipion pairs. Each $\Bs$ candidate is assigned to the PV with the smallest $\chisqip$, where \chisqip\ is defined as the difference in the vertex-fit \chisq to a given PV reconstructed with and without the candidate tracks being considered. A loose requirement that \chisqip\ is less than 25 effectively reduces combinatorial background. A kinematic vertex fit is applied to the $\Bs$ candidates. In this fit, to improve the mass resolution, the $\jpsi$ and $\etapr$ masses are constrained to their known values \cite{PDG2022}. The \chisq per degree of freedom of this fit is required to be less than 5. As discussed in Sec.~\ref{sec:Introduction}, only candidates with a decay time in the range $0.5-10 \ps$ are selected. As well as removing combinatorial background at low decay times, this defines a region where the acceptance is relatively flat.

 For both channels, specific vetoes are applied to remove background from exclusive $\bquark$-hadron decays. In the case of the $\Bs \to \jpsi \etapr$ decay mode, backgrounds from $B^0 \rightarrow \jpsi \pi^+ \pi^-$ and $B^0_s \rightarrow
 \jpsi \pi^+ \pi^-$ decays combined with a random photon are efficiently suppressed by rejecting  $\jpsi \pip \pim$ combinations with a reconstructed mass greater than 5249\mevcc. Background from the decay $\Bd \rightarrow \jpsi K^{*0} \left( \rightarrow K^+ \pi^- \right)$ is removed by tightening particle identification requirements if either of the two possible reconstructed $\jpsi \Kp \pim$ masses is within $30 \mevcc$ of the known $\Bd$ mass \cite{PDG2022}. Finally, background to the $\Bs \to \jpsi \etapr$ mode from the decay $\Bs \rightarrow \jpsi \phi(\rightarrow \pip \pim \piz )$ is suppressed by a requirement on the photon helicity angle.

 For the $\Bs \rightarrow \jpsi \pi^+ \pi^-$ decay mode, background from $\Bd \rightarrow \jpsi K^{*0} \left( \rightarrow K^+ \pi^- \right)$ is suppressed by the dipion mass window requirement and further reduced by the same veto used for the $\jpsi \etapr$ channel. Background from $B^+ \to J/ \psi h^+$ decays, where $h=\pi, K$, is suppressed by rejecting the candidate if any $\jpsi h^+$ combination has a mass within $\pm35 \mevcc$ of the known $\Bu$ mass \cite{PDG2022}.

The final step of the selection is to apply a multivariate classifier~\cite{Hocker:2007ht,*TMVA4} based on a gradient boosted decision tree (BDTG)~\cite{Roe_2005}. This classifier is trained to distinguish the simulated signal from the data in the background-dominated region where the B-candidate mass is between 5500 and 5650\mevcc. In the case of the $\Bs \to \jpsi \pip \pim$ decay mode, variables related to the $b$-hadron  kinematics, vertex quality and isolation, known to be compatible between data and simulation, are used. For the $\Bs \to \jpsi \etapr$ decay mode, four additional variables, related to kinematics and quality of the photon candidate, are added. The BDTG requirement for the $\Bs \to \jpsi \etapr$ mode is chosen, using pseudoexperiments, to minimise the uncertainty on $\Delta \Gamma_s$. In the case of the $\Bs \to \jpsi \pip \pim$ mode, since the decay only contains charged particles, the combinatorial background is lower and only a loose requirement on the BDTG response is needed. The chosen requirement keeps 90\% of the signal candidates while removing 95\% of the background. For both modes, the BDTG requirement does not bias the decay-time distribution. After the full selection, roughly 5\,\% of the events are found to contain more than one candidate. In this case, only one candidate chosen at random is kept.

\section{Invariant mass fit}
\label{sec:fitmodel}

\begin{figure}[htb!]
\begin{center}
\includegraphics[width=0.48\textwidth]{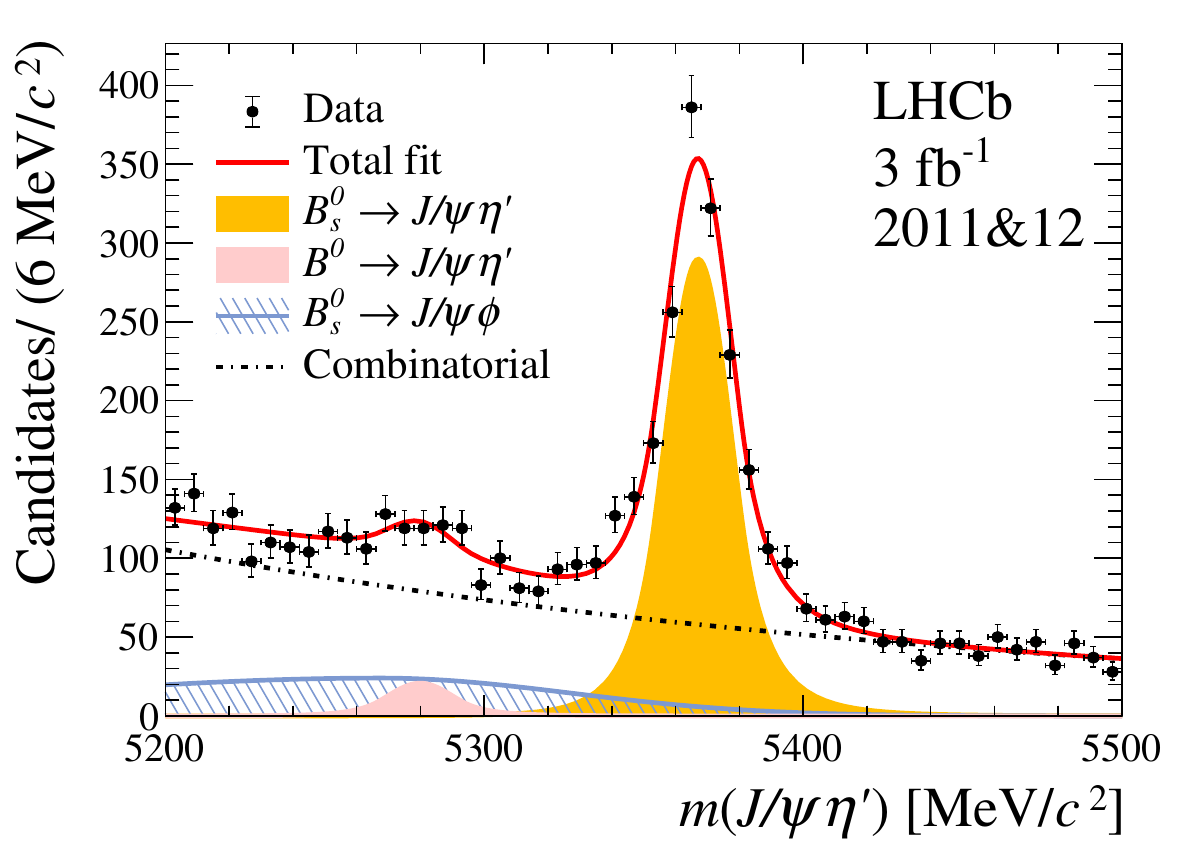}
\includegraphics[width=0.48\textwidth]{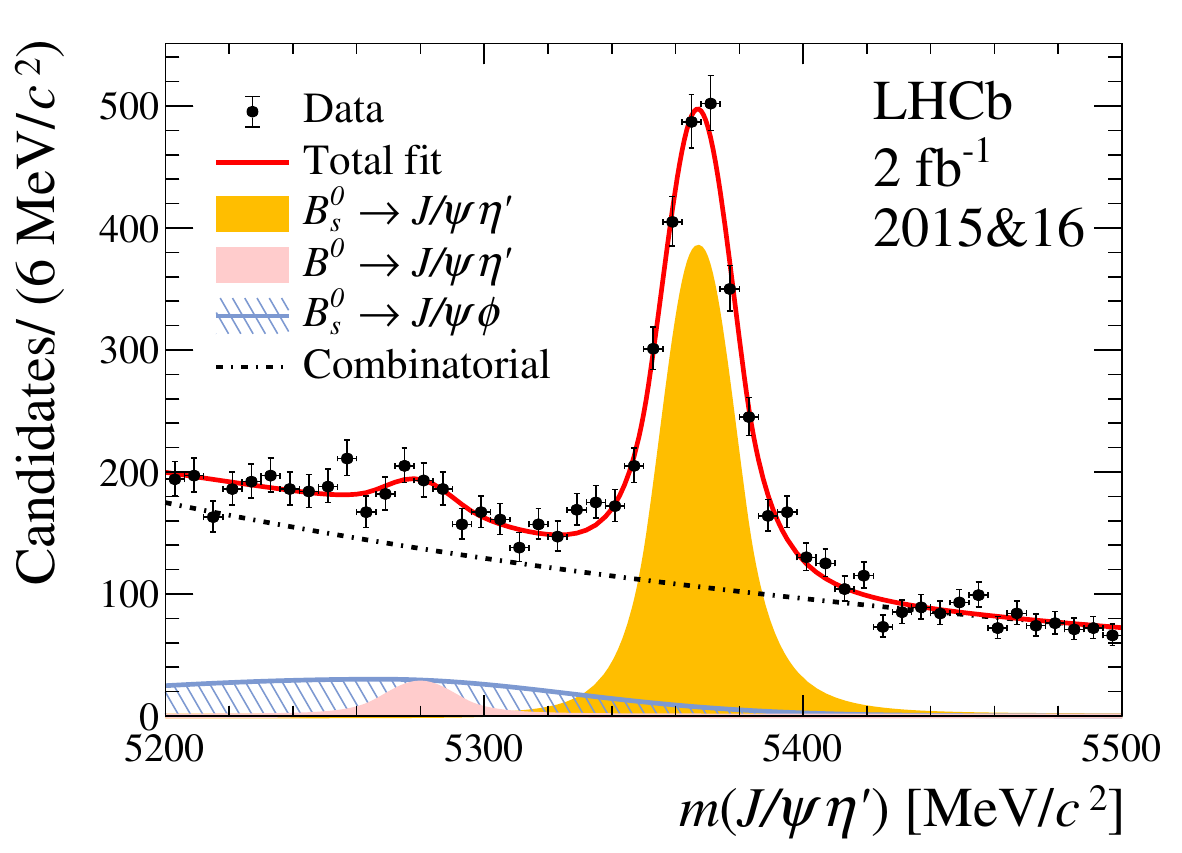}
\includegraphics[width=0.48\textwidth]{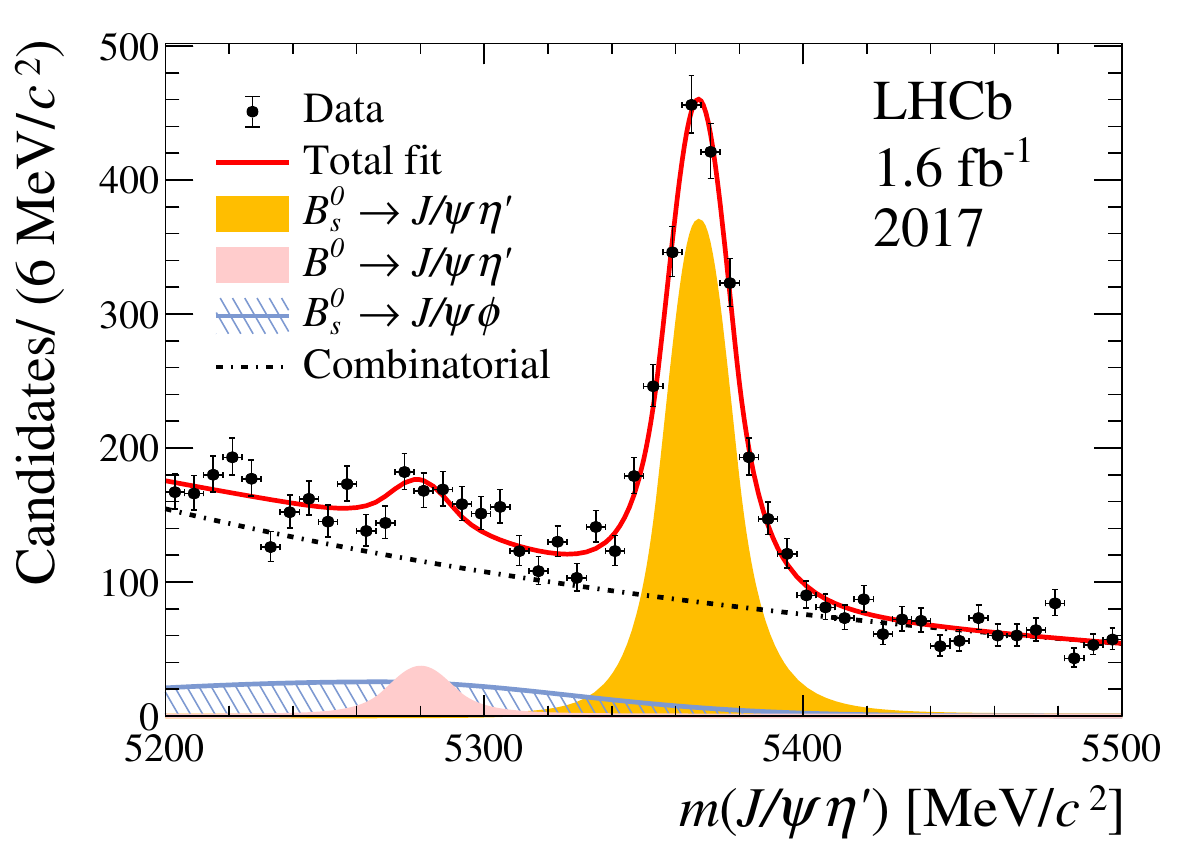}
\includegraphics[width=0.48\textwidth]{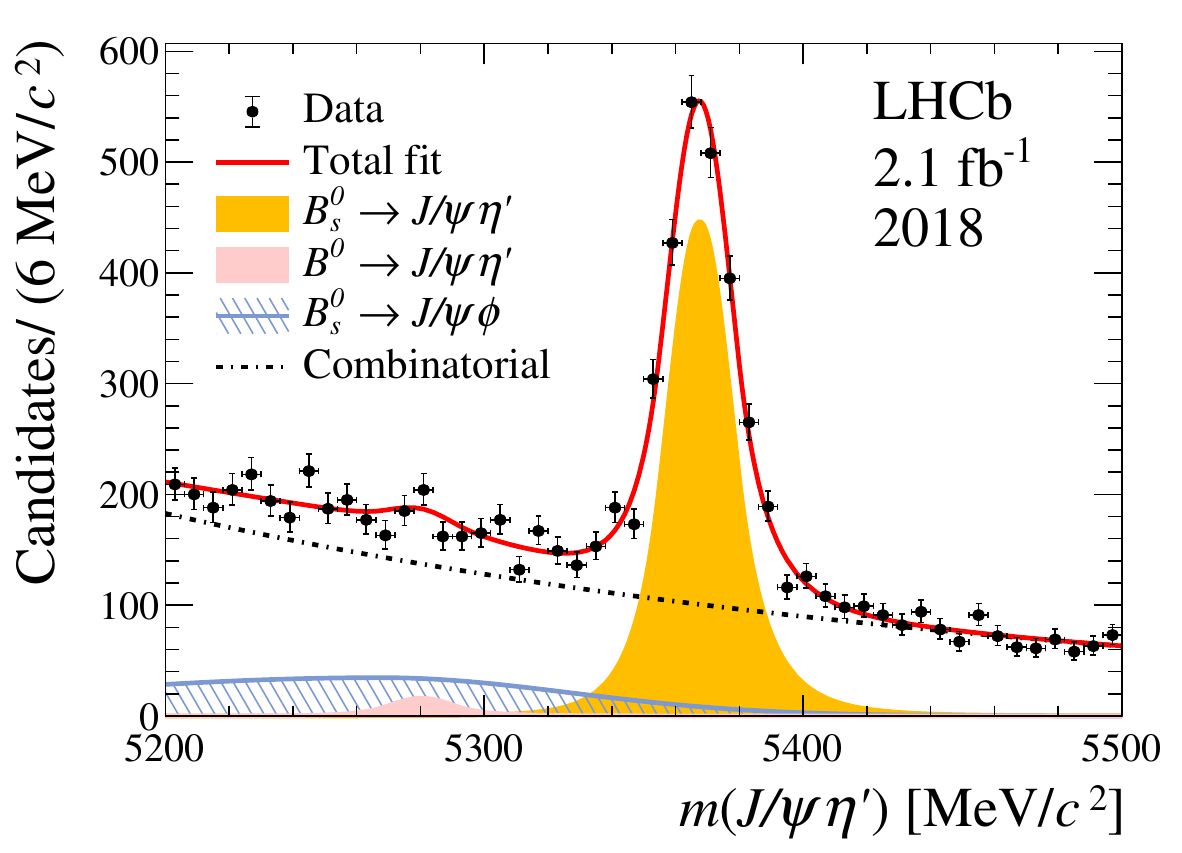}
\caption{\small Invariant mass distributions with the fit projection overlaid for the $\jpsi \etapr$ mode for the four datasets. For each dataset the eight time bins are summed.}
\label{fig:etaprsimfits}
\end{center}
\end{figure}

\begin{figure}[htb!]
\begin{center}
\includegraphics[width=0.48\textwidth]{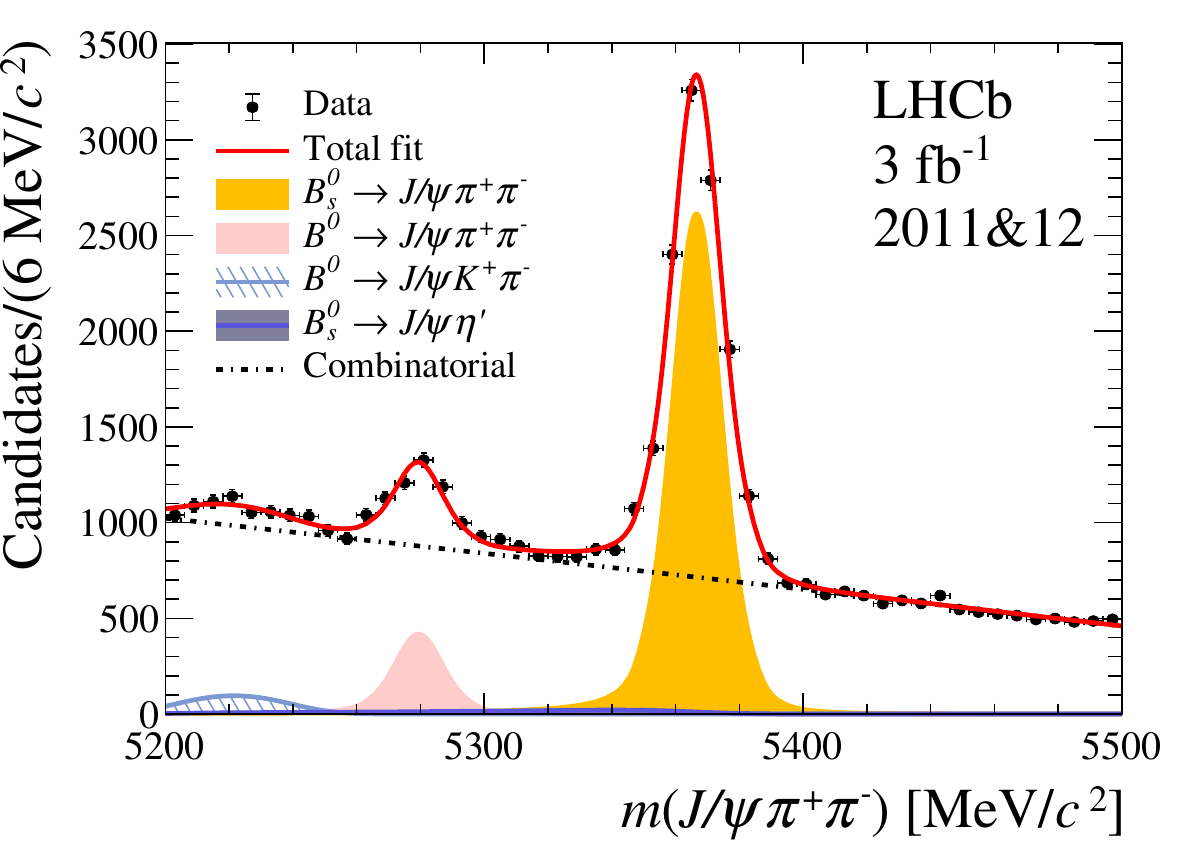}
\includegraphics[width=0.48\textwidth]{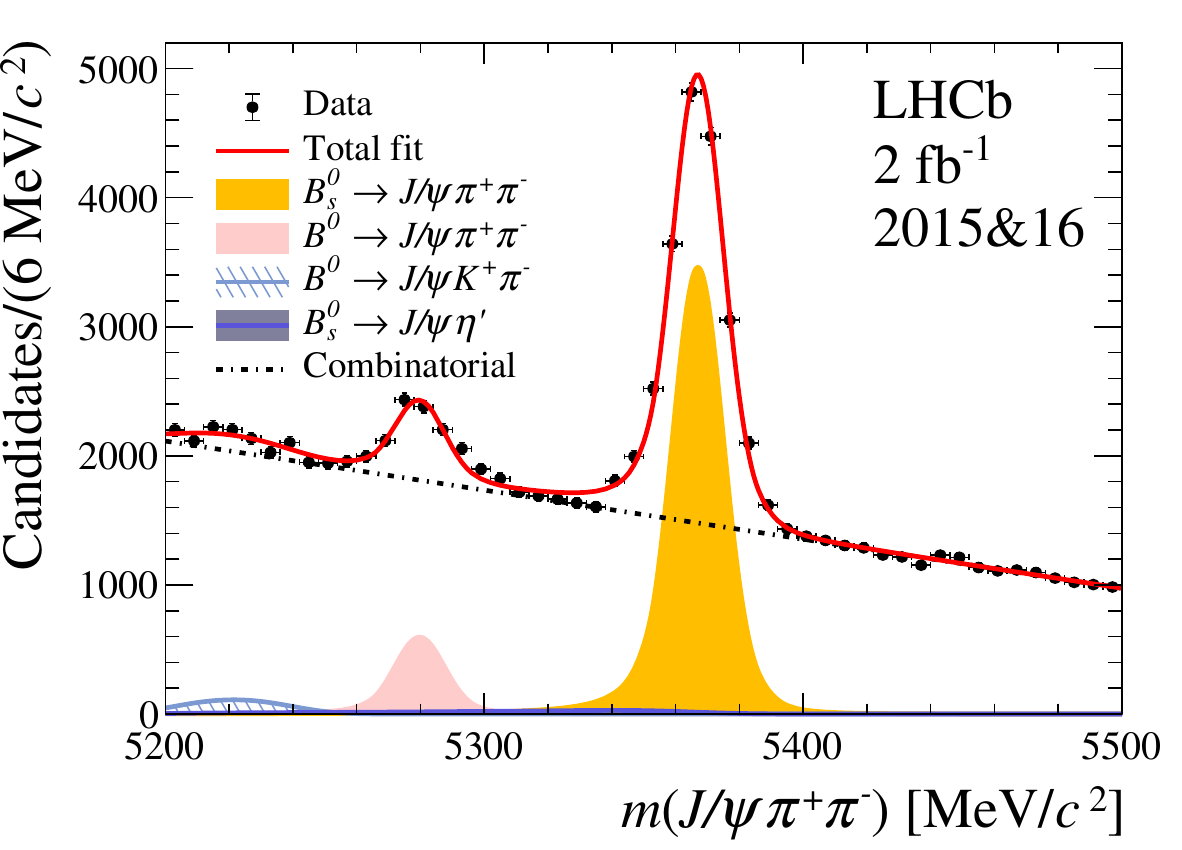}
\includegraphics[width=0.48\textwidth]{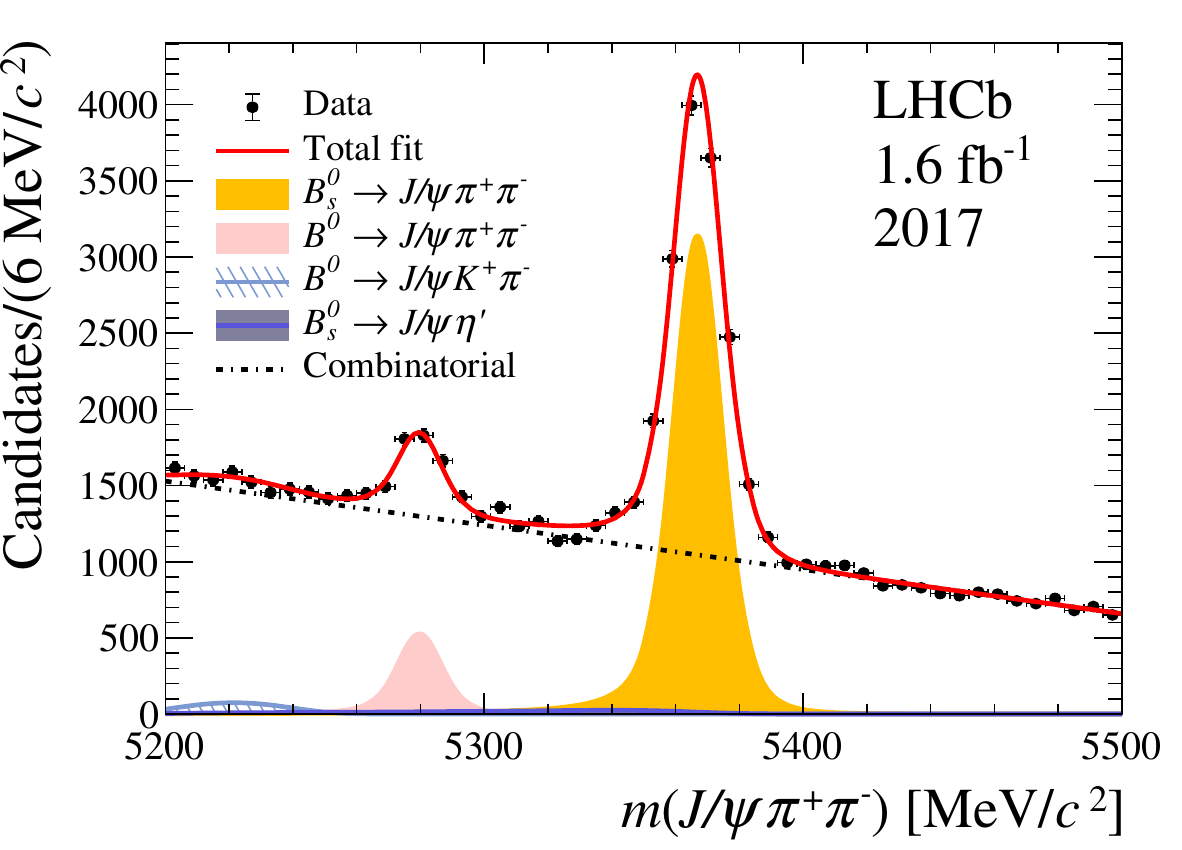}
\includegraphics[width=0.48\textwidth]{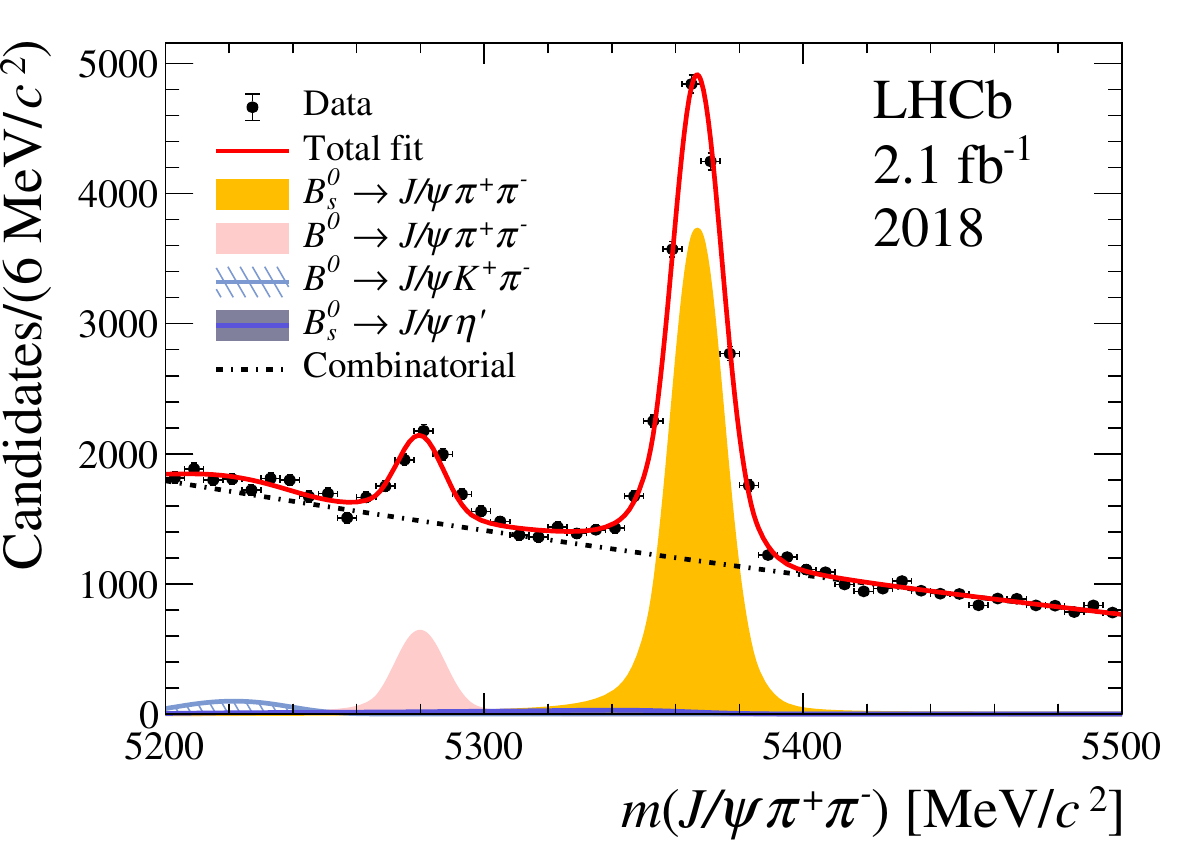}
\caption{\small Invariant mass distributions with the fit projection overlaid for the $\jpsi \pip \pim$ mode for the four datasets. For each dataset the eight time bins are summed.}
\label{fig:f0simfits}
\end{center}
\end{figure}

To obtain the yield in each decay-time bin, simultaneous extended unbinned maximum-likelihood fits are performed to the $\jpsi \etapr$ and the $\jpsi \pip \pim$ invariant mass distributions in the eight decay-time bins described in Sec.~\ref{sec:Introduction}. Fits are performed separately for the four datasets recorded in $2011$ and $2012, 2015$ and $2016, 2017, 2018$. The invariant mass distributions with the fit projection overlaid for each dataset are shown in Fig.~\ref{fig:etaprsimfits} for the $\Bs \to \jpsi \etapr$ candidates and in  Fig.~\ref{fig:f0simfits} for the $\Bs \to \jpsi \pip \pim$ candidates.

The fit model for the $J/\psi \etapr$ invariant-mass distribution has four components: the $\Bs \to \jpsi \etapr$  signal, the $\Bd \to \jpsi \etapr$ decay, the partially reconstructed decay $\Bs \to \jpsi \phi$  with $\phi \to \pi^+ \pi^- \pi^0$, and combinatorial background. The default signal model is chosen to be a double-sided Crystal Ball (DSCB) function. This is a generalization of the Crystal Ball function \cite{Skwarnicki:1986xj} with power law tails on both sides of the mass peak. In the fit to the data, the DSCB tail parameters are fixed to the values obtained from simulation, while the mean and resolution parameter ($\sigma_r$) are left free.

The $\Bd \to \jpsi \etapr$ decay is modelled using a DSCB shape with the same tail parameters as the $\Bs$ component. Since the mass resolution scales with the energy release (Q-value), the resolution parameter for this component is constrained to $s_Q \cdot \sigma_r$, where $s_Q = 0.97 \pm 0.02$. The difference in the positions of the $\Bs$ and $B^0$ mass peaks is also Gaussian-constrained to the known value of $m(\Bs) - m(\Bz) = 87.22 \pm 0.16 \mevcc$ \cite{PDG2022}. The yield of the \Bd component is left free in each decay-time bin since the fraction of $\Bs$ and $B^0$ decays depends on the decay-time.

 Partially reconstructed background from the $\Bs \to \jpsi \phi$ decay with $\phi \to \pi^+ \pi^- \pi^0$ is modelled with a bifurcated Gaussian function. In the fit to the data, the shape parameters are fixed to the values obtained from simulation while the relative fraction compared to the signal mode, $f_{\phi}$, is left free. The combinatorial background component is modelled by an exponential function, with the slope allowed to float independently in each of the eight decay-time bins.

The $\jpsi \pip \pim$ invariant-mass fit model has five components: the $\Bs \to \jpsi \pip \pim$ signal, the $\Bd \to \jpsi \pi^+ \pi^-$ decay, the misreconstructed $\Bd \to \jpsi K^+ \pi^-$ decay, $\Bs \to \jpsi \etapr$ decays with $\etapr \to \pi^+ \pi^- \gamma$, and combinatorial background. The invariant mass distribution of the $\Bs \to \jpsi \pip \pim$ decay is  well described by the sum of two DSCB functions with common mean and tail parameters. In the fit to the data, the tail parameters are fixed to the values obtained from simulation. Each of the two resolution parameters obtained from the simulation are multiplied by a scale-factor that varies freely in the fit. The $\Bd \to \jpsi \pi^+ \pi^-$ decay is dominated by an intermediate $\rho^0$ meson and is suppressed by the requirement that the dipion mass is within $90 \mevcc$ of the known mass of the $f_{0}(980)$ resonance. The remaining background from this source is modelled with the sum of two DSCB functions with the same tail and fraction parameters as for the $\Bs \rightarrow \jpsi \pip \pim$ decay. The position of the peak is Gaussian-constrained relative to the $\Bs$ and the mass resolution is constrained to the $\Bs$ mode assuming Q-value scaling.

The decay $B^0 \to \jpsi K^+ \pi^-$ is suppressed using the veto explained in Sec.~\ref{sec:selection}. The shape of the remaining background from this source is modelled using a histogram template obtained from the RapidSim simulation, including the convolution with the detector resolution. The size of this component is estimated by studying the $\jpsi K^+ \pim$ invariant mass distribution in data. Using these studies, the fraction of this component, whose uncertainty is dominated by the knowledge of the branching fraction, is constrained relative to the signal to be $0.01 \pm 0.1$.

Background from the partially reconstructed decay $\Bs \to \jpsi \etapr$ with $\etapr \to \pi^+ \pi^- \gamma$ is modelled using a histogram template from the RapidSim simulation, including the convolution with the detector resolution. The relative yield of the component to the signal decay is constrained to be $f_{\etapr}$= 0.6 $\pm$ 0.9\,\% using the known branching fractions and the relative efficiency obtained from the simulation. Finally, by combining pairs of same-sign pion candidates with selected $\jpsi$ candidates, the combinatorial background component is found to be well modelled by a second-order Chebychev polynomial. Based on the same-sign fits, in order to minimise the free parameters in the simultaneous fit to the data, the second-order parameter of this function is shared between the decay-time bins, while the first order parameter is left free and allowed to vary in each bin.

\section{Decay-time acceptance}
\label{sec:timeacc}
Due to the requirements made in the trigger and offline selection, the acceptance of the detector varies with decay time. At low decay times, the decay-length significance requirement removes events, while at high decay time, inefficiencies are introduced by the requirement on the candidate $\chisqip$ and the VELO track reconstruction \cite{LHCb-PAPER-2022-010}. The simulation is used to verify that the decay-time acceptance largely cancels due to the similarity of the two decays and their selections. The remaining relative acceptance correction is found to be well described by the form $A_{r}(t) = 1- \beta t$, using the simulation. The result of a fit to the simulation of this form is used to generate the acceptance correction, $A_i$, for each bin by evaluating $A_{r}(t)$ at the bin barycentre. The largest relative acceptance correction ($1.18$) is found in the last bin of the $2011$ and $2012$ dataset since the size of the VELO tracking-efficiency correction is large \cite{LHCb-PAPER-2013-065} for each mode, and thus it does not cancel in $A_i$. For the other datasets, improvements to the VELO tracking algorithm lead to a smaller relative acceptance correction of  $1.03$ or less. The overall effect of the relative acceptance correction is small. If it were ignored entirely, the central value of $\Delta \Gamma_s$ would change by half the statistical uncertainty. As part of the systematic studies, alternative functional forms and choices to evaluate the acceptance are considered. Another approach, used as a cross-check, is to assume the acceptance is flat within a decay-time bin. In this case, the relative acceptance is extracted from the simulation directly in the eight decay-time bins.

\section{Results and systematic uncertainties}
\label{sec:results}

\begin{figure}[htb!]
\begin{center}
\includegraphics[width=0.48\textwidth]{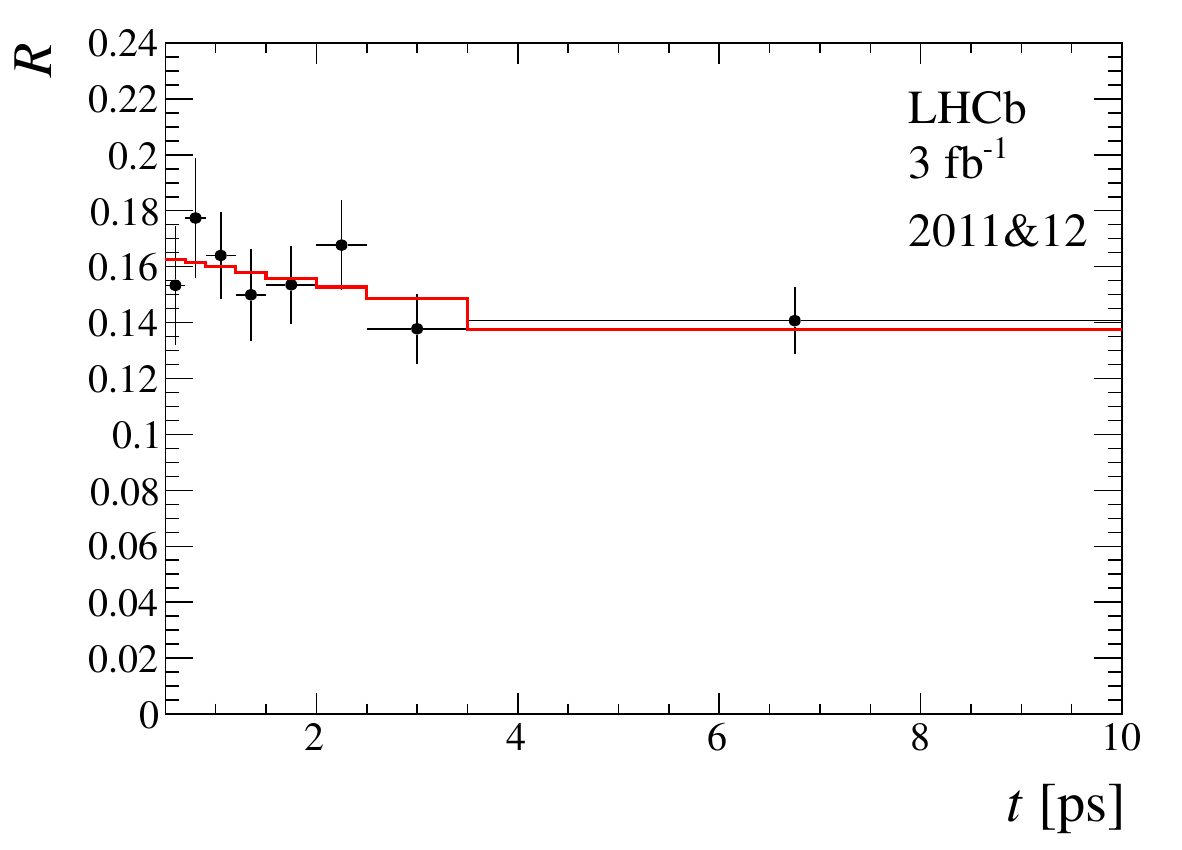}
\includegraphics[width=0.48\textwidth]{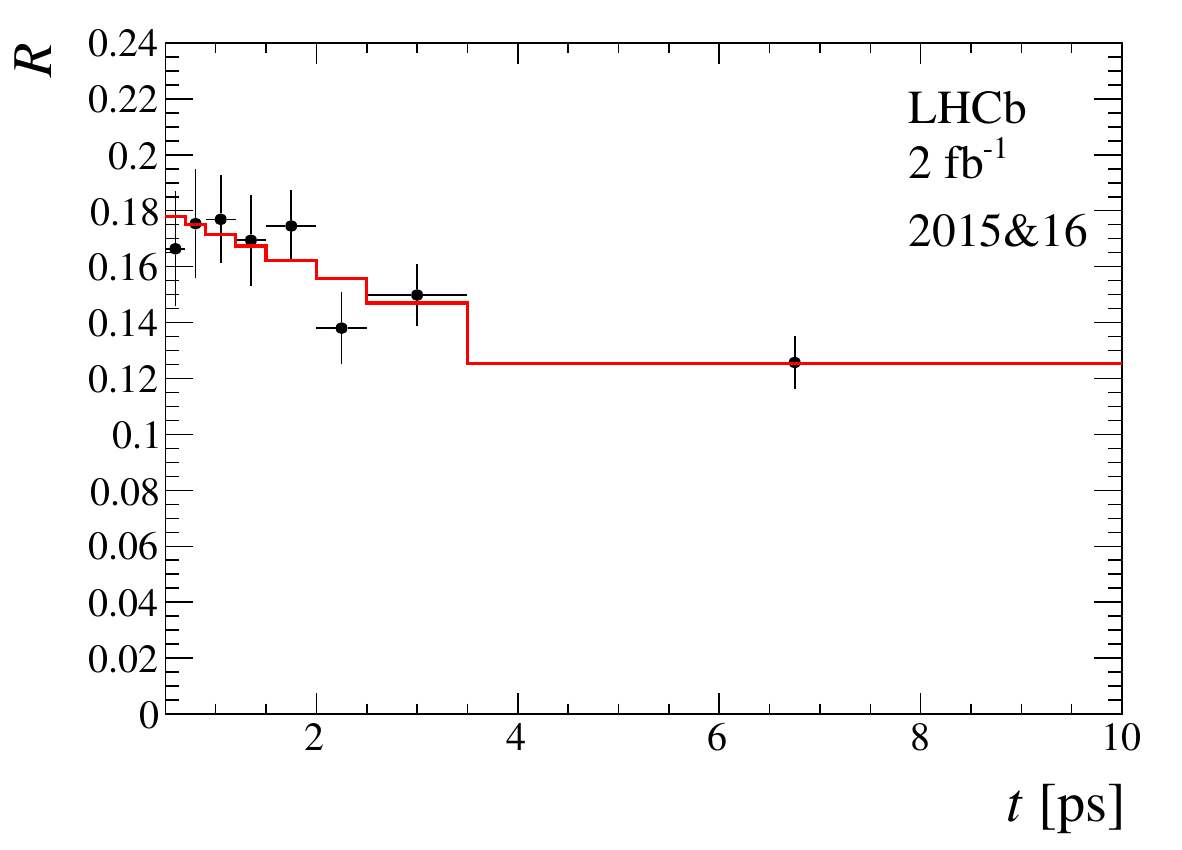}
\includegraphics[width=0.48\textwidth]{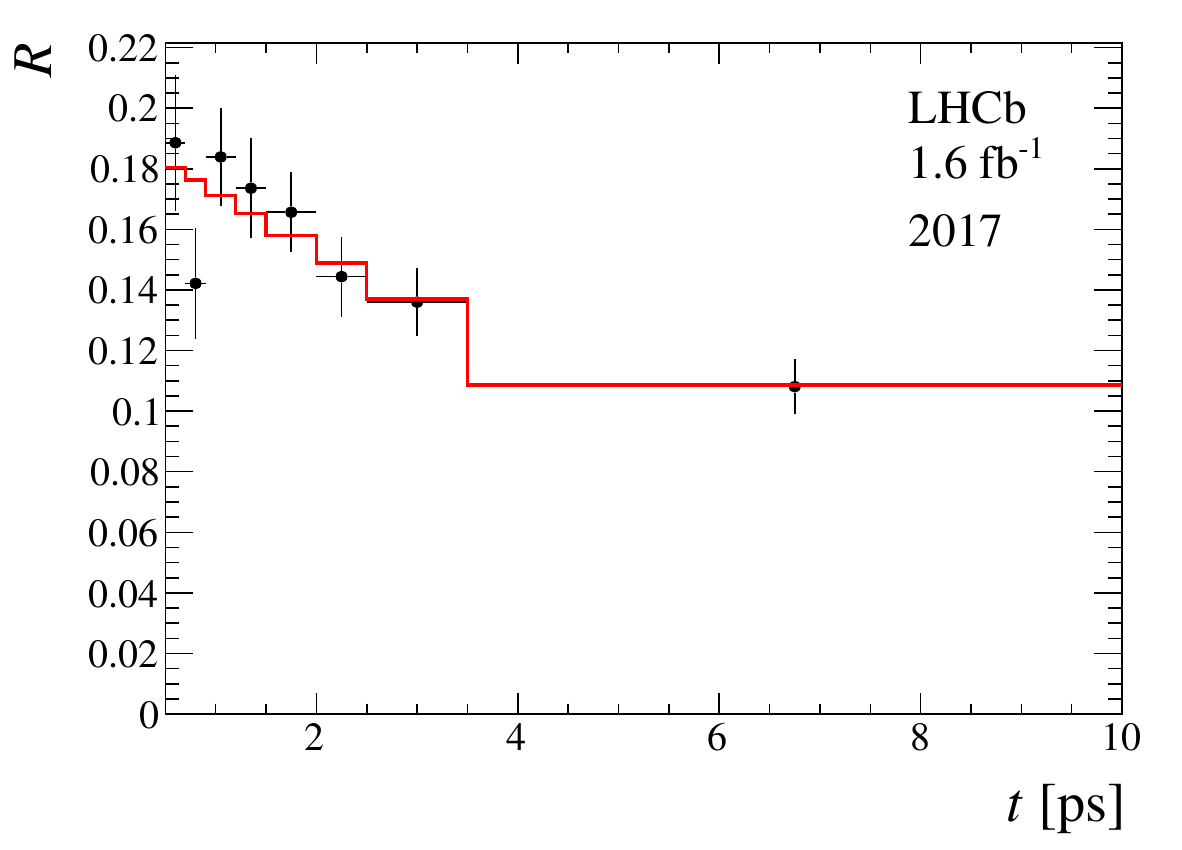}
\includegraphics[width=0.48\textwidth]{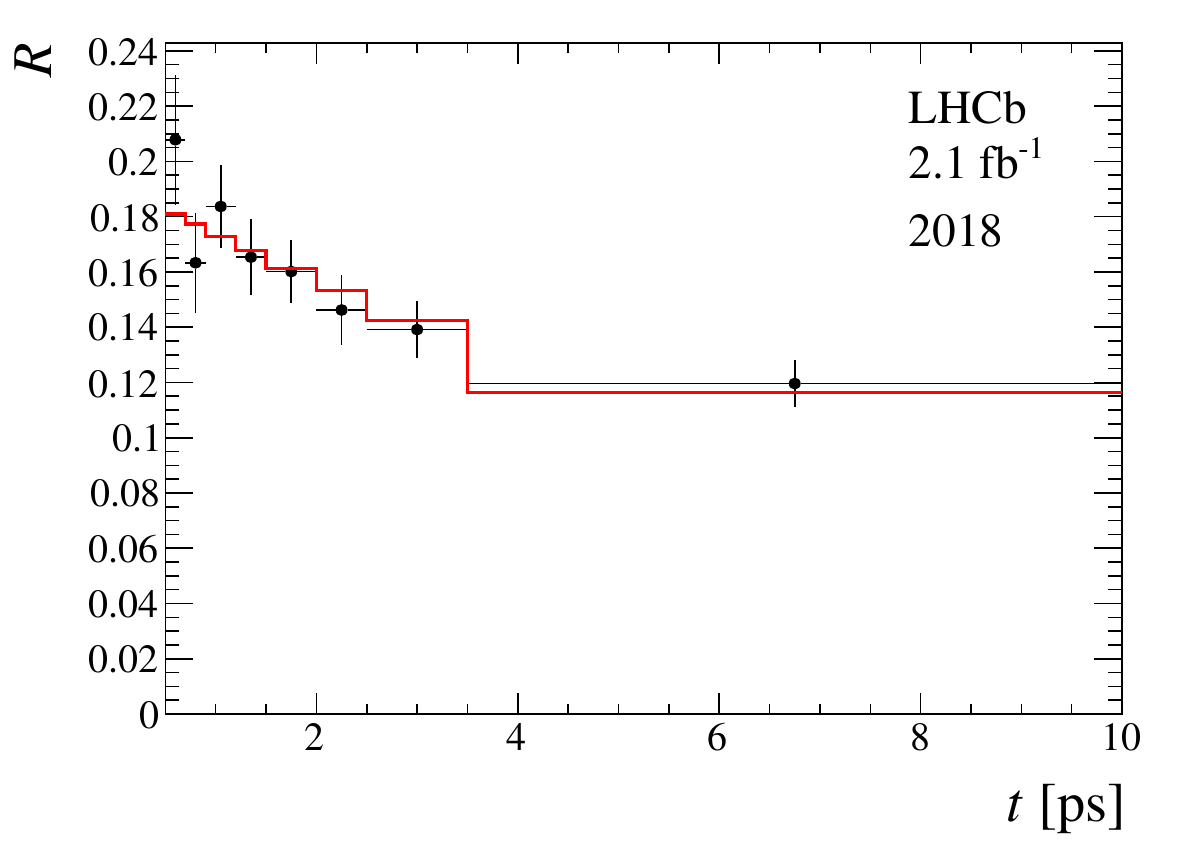}
\caption{Measurements of $R$ for the four datasets. The red line for each plot shows the result of the fit described in the text.}
\label{fig:dgresults}
\end{center}
\end{figure}

 The yields of the $\Bs \to \jpsi \etapr$ and $\Bs \to \jpsi \pip \pim$ decay are extracted from the extended unbinned simultaneous maximum-likelihood fit to the eight decay-time bins described in Sec.~\ref{sec:fitmodel}. The ratio of yields in each bin is corrected by the corresponding relative decay-time acceptance, and the \chisq fit described in Sec.~\ref{sec:Introduction} is performed. Figure~\ref{fig:dgresults} shows the fit result for each dataset. The resulting values of $\Delta \Gamma_s$ are summarized in Table~\ref{tab:results}. The weighted average of the results is $\Delta \Gamma_s = 0.087 \pm 0.012 \; \mathrm{ps}^{-1}$ where the uncertainty is statistical.

\begin{table}[]
    \centering
      \caption{\small Values of $\Delta \Gamma_s$ and the $\chi^2$ probability obtained from the fits to $R$ for the four datasets. The uncertainties are statistical.}
    \label{tab:results}
    \begin{tabular}{lcc}
        Dataset  & $\Delta \Gamma_s$ [$\ps^{-1}$] & P($\chi^2$) \\
        \hline
      2011$\&$12 & 0.039 $\pm$ 0.026 & 0.83 \\
      2015$\&$16 & 0.081 $\pm$ 0.022 & 0.77 \\
      2017 & 0.117 $\pm$ 0.024  & 0.57 \\
      2018 & 0.102 $\pm$ 0.021 & 0.78 \\
    \end{tabular}
\end{table}

Systematic uncertainties related to the knowledge of the detector acceptance largely cancel in $A_i$. The uncertainty from the finite size of the available simulation samples is estimated by resampling the covariance matrix of the fit used to obtain $\beta$, running the fit procedure and calculating the weighted average of the four datasets. This results in a systematic uncertainty of 4.6$\invns$. In the baseline fit, a linear model is used to determine $A_i$. To estimate the uncertainty due to this choice, a parabolic model and a histogram-based approach are considered. Based on these studies, an uncertainty of  $3.0 \invns$ is assigned. For the $2011$ and $2012$ dataset, where the acceptance corrections due to the trigger and selection are largest, additional data-driven checks of the acceptance ratio are made using the methods described in Ref.~\cite{LHCb-PAPER-2013-065}. Though the individual acceptance correction for each mode changes, $A_i$ is not significantly affected, and no further uncertainty is assigned.

The correctness of the fit procedure has been extensively verified using pseudoexperiments. The largest bias found in those tests arises from the choice of the decay-time value used to evaluate the acceptance correction. Based upon those studies, a $0.3 \invns$ uncertainty is assigned.

Several uncertainties arise from the limited knowledge of physics inputs to the fit. The method assumes that $\phi_s$ is zero. Experimentally, the current world average is $\phi_s = -0.049 \pm 0.019 \rad$ \cite{HFLAV21}. Based upon pseudoexperiments, a systematic uncertainty of  $0.1 \, \ns^{-1}$ is assigned to allow for a non-zero value of $\phi_s$. The analysis uses $\Bs \to \jpsi \pi^+ \pi^-$ decays with a dipion mass within $90 \mevcc$ of the known $f_0(980)$ mass, which are predominantly \CP-odd. The angular analysis in Ref.~\cite{LHCb-PAPER-2012-005}, limits the size of the \CP-even component in this dipion mass region to be less than 0.6\% at 95\% confidence level. Including a 0.6\% $\CP$-even component in the generation of pseudoexperiments and ignoring it in the fit results in a negligible bias. The default fit uses the current world average,  $\Gamma_{s} = 0.6628 \pm 0.0035 \, \ps^{-1}$ \cite{HFLAV21}, as input. Varying $\Gamma_{s}$ within its uncertainty results in a $0.1 \invns$ uncertainty on $\Delta \Gamma_s$.

Another source of uncertainty arises from the modelling of the signal and background components in the mass fits used to determine the yields. $\Delta \Gamma_{s}$ is measured using alternative models that describe the signal distribution and the results comparing to the measurement with the default fit model is found to be consistent. The impact of the background model is evaluated by varying the assumptions related to the modelling of the short-lived combinatorial and long-lived partially reconstructed components. The only significant variation is found to come from the combinatorial background model. For the $\Bs \rightarrow \jpsi \etapr$ mode, as an alternative to the
baseline exponential model, a first-order Chebychev function is considered. Using the fit results obtained for each bin, pseudoexperiments are generated with the default model, and then fit with the alternative background model. A bias of $6.9 \pm 0.1 \ns^{-1}$ is found for the $\Bs \to \jpsi \etapr$ mode and assigned as a systematic uncertainty. For the $\Bs \rightarrow \jpsi \pip \pim$ mode, generating pseudoexperiments with a first-order Chebychev polynomial and then fitting with a second-order (and vice versa) results in a bias of $0.8 \pm 0.1 \ns^{-1} $, which is assigned as the systematic uncertainty due to the background model.

\begin{figure}[]
    \centering
    \includegraphics[width=0.7\linewidth]{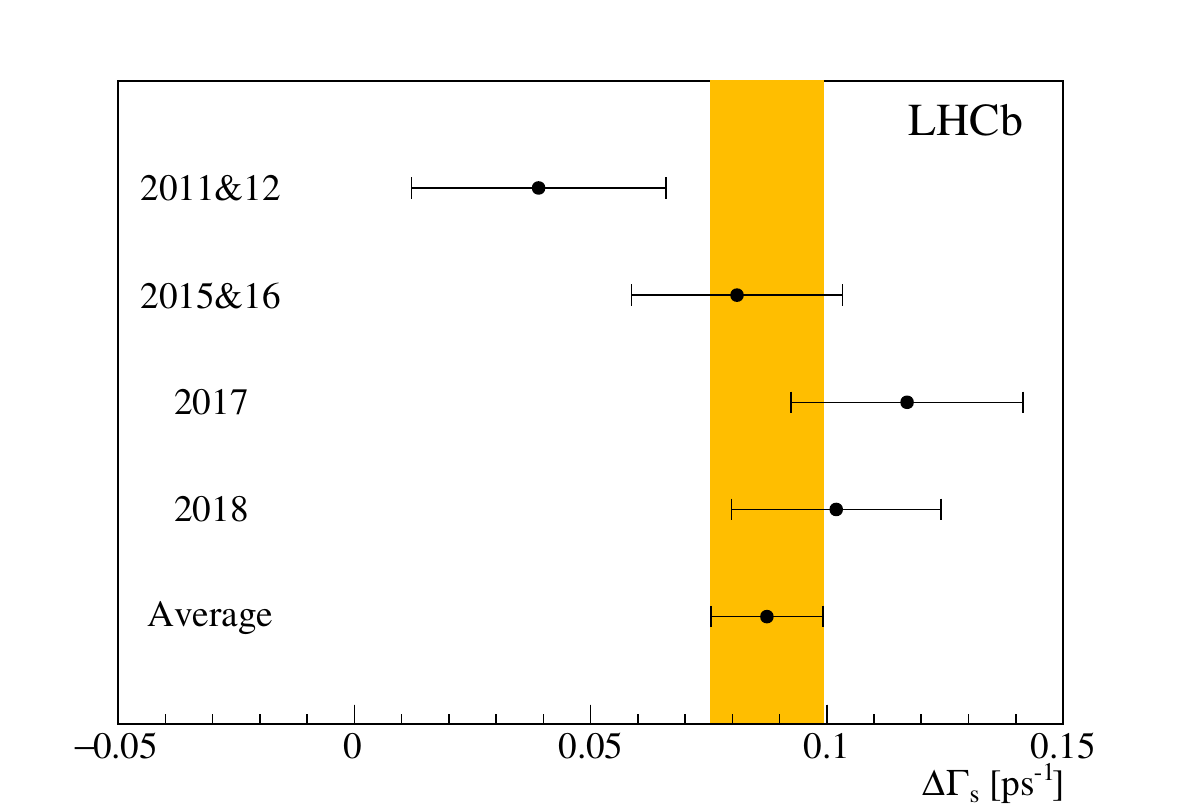}
    \caption[$\Delta \Gamma_s$ measurements.]{Measurement of $\Delta \Gamma_s$ for the four datasets and their weighted average.  The orange band is the $1\sigma$ error band.}
    \label{fig:delta_gammas_avg}
\end{figure}

The systematic uncertainties are summarized in Table~\ref{tab:systematics}. Adding them in quadrature leads to a total systematic uncertainty of $8.9 \invns$. The stability of the result is tested by comparing the results for the four datasets (Fig.~\ref{fig:delta_gammas_avg}). The $\chi^2$ probability for the four measurements, accounting for the statistical uncertainty and the uncorrelated part of the systematic uncertainty, is 12\,\% indicating that the four values are in good agreement.
\begin{table}[]
    \centering
      \caption{\small Systematic uncertainties on the measurement of $\Delta \Gamma_s$.}
    \label{tab:systematics}
    \begin{tabular}{lc}
        Source  & Value [$\ns^{-1}$]\\
        \hline
        Simulation sample size              & 4.6 \\
        Acceptance model                    & 3.0 \\
        Bin centre method                      & 0.3\\
        $\CP$ violation                      & 0.1\\
        $\Gamma_s$                          & 0.1\\
        $\jpsi \etapr$ background model       & 6.9\\
         $\jpsi \pip \pim$ background model       & 0.8 \\ \hline
        Total                     &  8.9 \\
    \end{tabular}
\end{table}

\section{Summary}
\label{sec:summary}
Using the full LHCb $pp$ collision dataset collected between 2011 and 2018, the $B^0_s \to \jpsi \eta^{\prime}$ and $B^0_s \to \jpsi \pi^{+} \pi^{-}$ decay modes are used to measure the decay-width difference
\begin{equation*}
\Delta \Gamma_s = 0.087 \pm 0.012 \pm 0.009  \; \mathrm{ps}^{-1},
\end{equation*}
where the first uncertainty is statistical and the second systematic.  This is the first time-dependent measurement using the $B^0_s \to \jpsi \eta^{\prime}$ decay mode. The value is in agreement with the HFLAV average,  $\Delta \Gamma_s = 0.074 \pm  0.006 \, \mathrm{ps}^{-1}$ \cite{HFLAV21}, determined from the time-dependent angular analyses of the $\Bs \to \jpsi \phi$ decay mode where the initial flavour of the state is tagged. It also agrees with the HFLAV average, $\Delta \Gamma_s = 0.083 \pm  0.005 \ps^{-1}$~\cite{HFLAV21}, which includes constraints from other untagged effective lifetime measurements. The alternative approach to determine $\Delta \Gamma_s$ presented in this paper may help to resolve the observed tensions between the measurements made by the LHC collaborations in the $\Bs \to \jpsi \phi$ mode.

\clearpage

\section*{Acknowledgements}
\noindent We express our gratitude to our colleagues in the CERN
accelerator departments for the excellent performance of the LHC. We
thank the technical and administrative staff at the LHCb
institutes.
We acknowledge support from CERN and from the national agencies:
CAPES, CNPq, FAPERJ and FINEP (Brazil);
MOST and NSFC (China);
CNRS/IN2P3 (France);
BMBF, DFG and MPG (Germany);
INFN (Italy);
NWO (Netherlands);
MNiSW and NCN (Poland);
MCID/IFA (Romania);
MICINN (Spain);
SNSF and SER (Switzerland);
NASU (Ukraine);
STFC (United Kingdom);
DOE NP and NSF (USA).
We acknowledge the computing resources that are provided by CERN, IN2P3
(France), KIT and DESY (Germany), INFN (Italy), SURF (Netherlands),
PIC (Spain), GridPP (United Kingdom),
CSCS (Switzerland), IFIN-HH (Romania), CBPF (Brazil),
Polish WLCG  (Poland) and NERSC (USA).
We are indebted to the communities behind the multiple open-source
software packages on which we depend.
Individual groups or members have received support from
ARC and ARDC (Australia);
Key Research Program of Frontier Sciences of CAS, CAS PIFI, CAS CCEPP,
Fundamental Research Funds for the Central Universities,
and Sci. \& Tech. Program of Guangzhou (China);
Minciencias (Colombia);
EPLANET, Marie Sk\l{}odowska-Curie Actions, ERC and NextGenerationEU (European Union);
A*MIDEX, ANR, IPhU and Labex P2IO, and R\'{e}gion Auvergne-Rh\^{o}ne-Alpes (France);
AvH Foundation (Germany);
ICSC (Italy);
GVA, XuntaGal, GENCAT, Inditex, InTalent and Prog.~Atracci\'on Talento, CM (Spain);
SRC (Sweden);
the Leverhulme Trust, the Royal Society
 and UKRI (United Kingdom).

\addcontentsline{toc}{section}{References}
\ifx\mcitethebibliography\mciteundefinedmacro
\PackageError{LHCb.bst}{mciteplus.sty has not been loaded}
{This bibstyle requires the use of the mciteplus package.}\fi
\providecommand{\href}[2]{#2}

\newpage

\centerline
{\large\bf LHCb collaboration}
\begin
{flushleft}
\small
R.~Aaij$^{35}$\lhcborcid{0000-0003-0533-1952},
A.S.W.~Abdelmotteleb$^{54}$\lhcborcid{0000-0001-7905-0542},
C.~Abellan~Beteta$^{48}$,
F.~Abudin{\'e}n$^{54}$\lhcborcid{0000-0002-6737-3528},
T.~Ackernley$^{58}$\lhcborcid{0000-0002-5951-3498},
B.~Adeva$^{44}$\lhcborcid{0000-0001-9756-3712},
M.~Adinolfi$^{52}$\lhcborcid{0000-0002-1326-1264},
P.~Adlarson$^{78}$\lhcborcid{0000-0001-6280-3851},
H.~Afsharnia$^{11}$,
C.~Agapopoulou$^{46}$\lhcborcid{0000-0002-2368-0147}
% ...

{\footnotesize \it

% ...
$^{11}$Universit{\'e} Clermont Auvergne, CNRS/IN2P3, LPC, Clermont-Ferrand, France\\
% ...
$^{35}$Nikhef National Institute for Subatomic Physics, Amsterdam, Netherlands\\
% ...
$^{44}$Instituto Galego de F{\'\i}sica de Altas Enerx{\'\i}as (IGFAE), Universidade de Santiago de Compostela, Santiago de Compostela, Spain\\
% ...
$^{46}$European Organization for Nuclear Research (CERN), Geneva, Switzerland\\
% ...
$^{48}$Physik-Institut, Universit{\"a}t Z{\"u}rich, Z{\"u}rich, Switzerland\\
% ...
$^{52}$H.H. Wills Physics Laboratory, University of Bristol, Bristol, United Kingdom\\
% ...
$^{54}$Department of Physics, University of Warwick, Coventry, United Kingdom\\
% ...
$^{57}$School of Physics and Astronomy, University of Glasgow, Glasgow, United Kingdom\\
$^{58}$Oliver Lodge Laboratory, University of Liverpool, Liverpool, United Kingdom\\
% ...
$^{78}$Department of Physics and Astronomy, Uppsala University, Uppsala, Sweden, associated to $^{57}$\\
% ...
}
\end{flushleft}


\begin{mcitethebibliography}{10}
\mciteSetBstSublistMode{n}
\mciteSetBstMaxWidthForm{subitem}{\alph{mcitesubitemcount})}
\mciteSetBstSublistLabelBeginEnd{\mcitemaxwidthsubitemform\space}
{\relax}{\relax}

\bibitem{Asatrian:2020zxa}
H.~M. Asatrian {\em et~al.}, \ifthenelse{\boolean{articletitles}}{\emph{Penguin
  contribution to the width difference and asymmetry in mixing},
  }{}\href{https://doi.org/10.1103/PhysRevD.102.033007}{Phys.\ Rev.\
  \textbf{D102} (2020) 033007},
  \href{http://arxiv.org/abs/2006.13227}{{\normalfont\ttfamily
  arXiv:2006.13227}}\relax
\mciteBstWouldAddEndPuncttrue
\mciteSetBstMidEndSepPunct{\mcitedefaultmidpunct}
{\mcitedefaultendpunct}{\mcitedefaultseppunct}\relax
\EndOfBibitem
\bibitem{Davies:2019gnp}
HPQCD collaboration, C.~T.~H. Davies {\em et~al.},
  \ifthenelse{\boolean{articletitles}}{\emph{Lattice {QCD} matrix elements for
  the ${B_s^0-\bar{B}_s^0}$ width difference},
  }{}\href{https://doi.org/10.1103/PhysRevLett.124.082001}{Phys.\ Rev.\ Lett.\
  \textbf{124} (2020) 082001},
  \href{http://arxiv.org/abs/1910.00970}{{\normalfont\ttfamily
  arXiv:1910.00970}}\relax
\mciteBstWouldAddEndPuncttrue
\mciteSetBstMidEndSepPunct{\mcitedefaultmidpunct}
{\mcitedefaultendpunct}{\mcitedefaultseppunct}\relax
\EndOfBibitem
\bibitem{Lenz:2019lvd}
A.~Lenz and G.~Tetlalmatzi-Xolocotzi,
  \ifthenelse{\boolean{articletitles}}{\emph{Model-independent bounds on new
  physics effects}, }{}\href{https://doi.org/10.1007/JHEP07(2020)177}{JHEP
  \textbf{07} (2020) 177},
  \href{http://arxiv.org/abs/1912.07621}{{\normalfont\ttfamily
  arXiv:1912.07621}}\relax
\mciteBstWouldAddEndPuncttrue
\mciteSetBstMidEndSepPunct{\mcitedefaultmidpunct}
{\mcitedefaultendpunct}{\mcitedefaultseppunct}\relax
\EndOfBibitem
\bibitem{Gerlach:2022wgb}
M.~Gerlach {\em et~al.}, \ifthenelse{\boolean{articletitles}}{\emph{{The width
  difference in $B - \bar B$ mixing at order $\alpha_s$ and beyond}},
  }{}\href{https://doi.org/10.1007/JHEP04(2022)006}{JHEP \textbf{04} (2022)
  006}, \href{http://arxiv.org/abs/2202.12305}{{\normalfont\ttfamily
  arXiv:2202.12305}}\relax
\mciteBstWouldAddEndPuncttrue
\mciteSetBstMidEndSepPunct{\mcitedefaultmidpunct}
{\mcitedefaultendpunct}{\mcitedefaultseppunct}\relax
\EndOfBibitem
\bibitem{ATLAS:2020lbz}
ATLAS collaboration, G.~Aad {\em et~al.},
  \ifthenelse{\boolean{articletitles}}{\emph{{Measurement of the $CP$-violating
  phase $\phi_s$ in $B^0_s \to J/\psi\phi$ decays in ATLAS at 13 TeV}},
  }{}\href{https://doi.org/10.1140/epjc/s10052-021-09011-0}{Eur.\ Phys.\ J.\
  \textbf{C81} (2021) 342},
  \href{http://arxiv.org/abs/2001.07115}{{\normalfont\ttfamily
  arXiv:2001.07115}}\relax
\mciteBstWouldAddEndPuncttrue
\mciteSetBstMidEndSepPunct{\mcitedefaultmidpunct}
{\mcitedefaultendpunct}{\mcitedefaultseppunct}\relax
\EndOfBibitem
\bibitem{CMS:2020efq}
CMS collaboration, A.~M. Sirunyan {\em et~al.},
  \ifthenelse{\boolean{articletitles}}{\emph{{Measurement of the $CP$-violating
  phase $\phi_\mathrm{s}$ in the B$^0_\mathrm{s}\to$ J$/\psi\, \phi$ $\to
  \mu^+\mu^-$K$^+$K$^-$ channel in proton-proton collisions at $\sqrt{s} =$ 13
  TeV}}, }{}\href{https://doi.org/10.1016/j.physletb.2021.136188}{Phys.\ Lett.\
   \textbf{B816} (2021) 136188},
  \href{http://arxiv.org/abs/2007.02434}{{\normalfont\ttfamily
  arXiv:2007.02434}}\relax
\mciteBstWouldAddEndPuncttrue
\mciteSetBstMidEndSepPunct{\mcitedefaultmidpunct}
{\mcitedefaultendpunct}{\mcitedefaultseppunct}\relax
\EndOfBibitem
\bibitem{LHCb-PAPER-2023-016}
LHCb collaboration, R.~Aaij {\em et~al.},
  \ifthenelse{\boolean{articletitles}}{\emph{{Improved measurement of CP
  violation parameters in $\Bs\to \jpsi \Kp\Km$ decays in the vicinty of the
  $\phiz(1020)$ resonance}},
  }{}\href{http://arxiv.org/abs/2308.01468}{{\normalfont\ttfamily
  arXiv:2308.01468}}, {submitted to Phys. Rev. Lett.}\relax
\mciteBstWouldAddEndPunctfalse
\mciteSetBstMidEndSepPunct{\mcitedefaultmidpunct}
{}{\mcitedefaultseppunct}\relax
\EndOfBibitem
\bibitem{HFLAV21}
Y.~Amhis {\em et~al.}, \ifthenelse{\boolean{articletitles}}{\emph{{Averages of
  $b$-hadron, $c$-hadron, and $\tau$-lepton properties as of 2021}},
  }{}\href{https://doi.org/10.1103/PhysRevD.107.052008}{Phys.\ Rev.\
  \textbf{D107} (2023) 052008},
  \href{http://arxiv.org/abs/2206.07501}{{\normalfont\ttfamily
  arXiv:2206.07501}}, {updated results and plots available at
  \href{https://hflav.web.cern.ch}{{\texttt{https://hflav.web.cern.ch}}}}\relax
\mciteBstWouldAddEndPuncttrue
\mciteSetBstMidEndSepPunct{\mcitedefaultmidpunct}
{\mcitedefaultendpunct}{\mcitedefaultseppunct}\relax
\EndOfBibitem
\bibitem{LHCB-PAPER-2017-008}
LHCb collaboration, R.~Aaij {\em et~al.},
  \ifthenelse{\boolean{articletitles}}{\emph{{Resonances and \CP-violation in
  \Bs and \mbox{\decay{\Bsb}{\jpsi\Kp\Km}} decays in the mass region above the
  $\phiz(1020)$}}, }{}\href{https://doi.org/10.1007/JHEP08(2017)037}{JHEP
  \textbf{08} (2017) 037},
  \href{http://arxiv.org/abs/1704.08217}{{\normalfont\ttfamily
  arXiv:1704.08217}}\relax
\mciteBstWouldAddEndPuncttrue
\mciteSetBstMidEndSepPunct{\mcitedefaultmidpunct}
{\mcitedefaultendpunct}{\mcitedefaultseppunct}\relax
\EndOfBibitem
\bibitem{LHCB-PAPER-2016-027}
LHCb collaboration, R.~Aaij {\em et~al.},
  \ifthenelse{\boolean{articletitles}}{\emph{{Measurement of the \CP violating
  phase and decay-width difference in \mbox{\decay{\Bs}{\psitwos\phi}}
  decays}}, }{}\href{https://doi.org/10.1016/j.physletb.2016.09.028}{Phys.\
  Lett.\  \textbf{B762} (2016) 253},
  \href{http://arxiv.org/abs/1608.04855}{{\normalfont\ttfamily
  arXiv:1608.04855}}\relax
\mciteBstWouldAddEndPuncttrue
\mciteSetBstMidEndSepPunct{\mcitedefaultmidpunct}
{\mcitedefaultendpunct}{\mcitedefaultseppunct}\relax
\EndOfBibitem
\bibitem{Fleischer:2011cw}
R.~Fleischer and R.~Knegjens,
  \ifthenelse{\boolean{articletitles}}{\emph{{Effective lifetime of $B_s$
  decays and $B_s^0$-$\bar B_s^0$ mixing parameters}},
  }{}\href{https://doi.org/{10.1140/epjc/s10052-011-1789-9}}{Eur.\ Phys.\ J.\
  \textbf{C71} (2011) 1789},
  \href{http://arxiv.org/abs/1109.5115}{{\normalfont\ttfamily
  arXiv:1109.5115}}\relax
\mciteBstWouldAddEndPuncttrue
\mciteSetBstMidEndSepPunct{\mcitedefaultmidpunct}
{\mcitedefaultendpunct}{\mcitedefaultseppunct}\relax
\EndOfBibitem
\bibitem{PDG2022}
Particle Data Group, R.~L. Workman {\em et~al.},
  \ifthenelse{\boolean{articletitles}}{\emph{{\href{http://pdg.lbl.gov/}{Review
  of particle physics}}}, }{}\href{https://doi.org/10.1093/ptep/ptac097}{Prog.\
  Theor.\ Exp.\ Phys.\  \textbf{2022} (2022) 083C01}\relax
\mciteBstWouldAddEndPuncttrue
\mciteSetBstMidEndSepPunct{\mcitedefaultmidpunct}
{\mcitedefaultendpunct}{\mcitedefaultseppunct}\relax
\EndOfBibitem
\bibitem{LHCb-PAPER-2013-069}
LHCb collaboration, R.~Aaij {\em et~al.},
  \ifthenelse{\boolean{articletitles}}{\emph{{Measurement of resonant and \CP
  components in \mbox{\decay{\Bsb}{\jpsi\pip\pim}} decays}},
  }{}\href{https://doi.org/10.1103/PhysRevD.89.092006}{Phys.\ Rev.\
  \textbf{D89} (2014) 092006},
  \href{http://arxiv.org/abs/1402.6248}{{\normalfont\ttfamily
  arXiv:1402.6248}}\relax
\mciteBstWouldAddEndPuncttrue
\mciteSetBstMidEndSepPunct{\mcitedefaultmidpunct}
{\mcitedefaultendpunct}{\mcitedefaultseppunct}\relax
\EndOfBibitem
\bibitem{Aaboud:2016bro}
ATLAS collaboration, M.~Aaboud {\em et~al.},
  \ifthenelse{\boolean{articletitles}}{\emph{{Measurement of the relative width
  difference of the $B^0$-$\bar B^0$ system with the ATLAS detector}},
  }{}\href{https://doi.org/10.1007/JHEP06(2016)081}{JHEP \textbf{06} (2016)
  081}, \href{http://arxiv.org/abs/1605.07485}{{\normalfont\ttfamily
  arXiv:1605.07485}}\relax
\mciteBstWouldAddEndPuncttrue
\mciteSetBstMidEndSepPunct{\mcitedefaultmidpunct}
{\mcitedefaultendpunct}{\mcitedefaultseppunct}\relax
\EndOfBibitem
\bibitem{Lafferty:1994cj}
G.~D. Lafferty and T.~R. Wyatt,
  \ifthenelse{\boolean{articletitles}}{\emph{Where to stick your data points:
  The treatment of measurements within wide bins},
  }{}\href{https://doi.org/10.1016/0168-9002(94)01112-5}{Nucl.\ Instrum.\
  Meth.\  \textbf{A355} (1995) 541}\relax
\mciteBstWouldAddEndPuncttrue
\mciteSetBstMidEndSepPunct{\mcitedefaultmidpunct}
{\mcitedefaultendpunct}{\mcitedefaultseppunct}\relax
\EndOfBibitem
\bibitem{LHCb-DP-2012-002}
A.~A. Alves~Jr.\ {\em et~al.},
  \ifthenelse{\boolean{articletitles}}{\emph{{Performance of the LHCb muon
  system}}, }{}\href{https://doi.org/10.1088/1748-0221/8/02/P02022}{JINST
  \textbf{8} (2013) P02022},
  \href{http://arxiv.org/abs/1211.1346}{{\normalfont\ttfamily
  arXiv:1211.1346}}\relax
\mciteBstWouldAddEndPuncttrue
\mciteSetBstMidEndSepPunct{\mcitedefaultmidpunct}
{\mcitedefaultendpunct}{\mcitedefaultseppunct}\relax
\EndOfBibitem
\bibitem{LHCb-DP-2014-002}
LHCb collaboration, R.~Aaij {\em et~al.},
  \ifthenelse{\boolean{articletitles}}{\emph{{LHCb detector performance}},
  }{}\href{https://doi.org/10.1142/S0217751X15300227}{Int.\ J.\ Mod.\ Phys.\
  \textbf{A30} (2015) 1530022},
  \href{http://arxiv.org/abs/1412.6352}{{\normalfont\ttfamily
  arXiv:1412.6352}}\relax
\mciteBstWouldAddEndPuncttrue
\mciteSetBstMidEndSepPunct{\mcitedefaultmidpunct}
{\mcitedefaultendpunct}{\mcitedefaultseppunct}\relax
\EndOfBibitem
\bibitem{LHCb-PAPER-2013-011}
LHCb collaboration, R.~Aaij {\em et~al.},
  \ifthenelse{\boolean{articletitles}}{\emph{{Precision measurement of \D meson
  mass differences}}, }{}\href{https://doi.org/10.1007/JHEP06(2013)065}{JHEP
  \textbf{06} (2013) 065},
  \href{http://arxiv.org/abs/1304.6865}{{\normalfont\ttfamily
  arXiv:1304.6865}}\relax
\mciteBstWouldAddEndPuncttrue
\mciteSetBstMidEndSepPunct{\mcitedefaultmidpunct}
{\mcitedefaultendpunct}{\mcitedefaultseppunct}\relax
\EndOfBibitem
\bibitem{LHCb-DP-2020-001}
C.~Abellan~Beteta {\em et~al.},
  \ifthenelse{\boolean{articletitles}}{\emph{{Calibration and performance of
  the LHCb calorimeters in Run 1 and 2 at the LHC}},
  }{}\href{http://arxiv.org/abs/2008.11556}{{\normalfont\ttfamily
  arXiv:2008.11556}}, {submitted to JINST}\relax
\mciteBstWouldAddEndPuncttrue
\mciteSetBstMidEndSepPunct{\mcitedefaultmidpunct}
{\mcitedefaultendpunct}{\mcitedefaultseppunct}\relax
\EndOfBibitem
\bibitem{LHCb-DP-2019-001}
R.~Aaij {\em et~al.}, \ifthenelse{\boolean{articletitles}}{\emph{{Performance
  of the LHCb trigger and full real-time reconstruction in Run 2 of the LHC}},
  }{}\href{https://doi.org/10.1088/1748-0221/14/04/P04013}{JINST \textbf{14}
  (2019) P04013}, \href{http://arxiv.org/abs/1812.10790}{{\normalfont\ttfamily
  arXiv:1812.10790}}\relax
\mciteBstWouldAddEndPuncttrue
\mciteSetBstMidEndSepPunct{\mcitedefaultmidpunct}
{\mcitedefaultendpunct}{\mcitedefaultseppunct}\relax
\EndOfBibitem
\bibitem{Borghi:2017hfp}
S.~Borghi, \ifthenelse{\boolean{articletitles}}{\emph{{Novel real-time
  alignment and calibration of the LHCb detector and its performance}},
  }{}\href{https://doi.org/10.1016/j.nima.2016.06.050}{Nucl.\ Instrum.\ Meth.\
  \textbf{A845} (2017) 560}\relax
\mciteBstWouldAddEndPuncttrue
\mciteSetBstMidEndSepPunct{\mcitedefaultmidpunct}
{\mcitedefaultendpunct}{\mcitedefaultseppunct}\relax
\EndOfBibitem
\bibitem{Sjostrand:2007gs}
T.~Sj\"{o}strand, S.~Mrenna, and P.~Skands,
  \ifthenelse{\boolean{articletitles}}{\emph{{A brief introduction to PYTHIA
  8.1}}, }{}\href{https://doi.org/10.1016/j.cpc.2008.01.036}{Comput.\ Phys.\
  Commun.\  \textbf{178} (2008) 852},
  \href{http://arxiv.org/abs/0710.3820}{{\normalfont\ttfamily
  arXiv:0710.3820}}\relax
\mciteBstWouldAddEndPuncttrue
\mciteSetBstMidEndSepPunct{\mcitedefaultmidpunct}
{\mcitedefaultendpunct}{\mcitedefaultseppunct}\relax
\EndOfBibitem
\bibitem{LHCb-PROC-2010-056}
I.~Belyaev {\em et~al.}, \ifthenelse{\boolean{articletitles}}{\emph{{Handling
  of the generation of primary events in Gauss, the LHCb simulation
  framework}}, }{}\href{https://doi.org/10.1088/1742-6596/331/3/032047}{J.\
  Phys.\ Conf.\ Ser.\  \textbf{331} (2011) 032047}\relax
\mciteBstWouldAddEndPuncttrue
\mciteSetBstMidEndSepPunct{\mcitedefaultmidpunct}
{\mcitedefaultendpunct}{\mcitedefaultseppunct}\relax
\EndOfBibitem
\bibitem{Lange:2001uf}
D.~J. Lange, \ifthenelse{\boolean{articletitles}}{\emph{{The EvtGen particle
  decay simulation package}},
  }{}\href{https://doi.org/10.1016/S0168-9002(01)00089-4}{Nucl.\ Instrum.\
  Meth.\  \textbf{A462} (2001) 152}\relax
\mciteBstWouldAddEndPuncttrue
\mciteSetBstMidEndSepPunct{\mcitedefaultmidpunct}
{\mcitedefaultendpunct}{\mcitedefaultseppunct}\relax
\EndOfBibitem
\bibitem{Golonka:2005pn}
P.~Golonka and Z.~Was, \ifthenelse{\boolean{articletitles}}{\emph{{PHOTOS Monte
  Carlo: A precision tool for QED corrections in $Z$ and $W$ decays}},
  }{}\href{https://doi.org/10.1140/epjc/s2005-02396-4}{Eur.\ Phys.\ J.\
  \textbf{C45} (2006) 97},
  \href{http://arxiv.org/abs/hep-ph/0506026}{{\normalfont\ttfamily
  arXiv:hep-ph/0506026}}\relax
\mciteBstWouldAddEndPuncttrue
\mciteSetBstMidEndSepPunct{\mcitedefaultmidpunct}
{\mcitedefaultendpunct}{\mcitedefaultseppunct}\relax
\EndOfBibitem
\bibitem{Allison:2006ve}
Geant4 collaboration, J.~Allison {\em et~al.},
  \ifthenelse{\boolean{articletitles}}{\emph{{Geant4 developments and
  applications}}, }{}\href{https://doi.org/10.1109/TNS.2006.869826}{IEEE
  Trans.\ Nucl.\ Sci.\  \textbf{53} (2006) 270}\relax
\mciteBstWouldAddEndPuncttrue
\mciteSetBstMidEndSepPunct{\mcitedefaultmidpunct}
{\mcitedefaultendpunct}{\mcitedefaultseppunct}\relax
\EndOfBibitem
\bibitem{Agostinelli:2002hh}
Geant4 collaboration, S.~Agostinelli {\em et~al.},
  \ifthenelse{\boolean{articletitles}}{\emph{{Geant4: A simulation toolkit}},
  }{}\href{https://doi.org/10.1016/S0168-9002(03)01368-8}{Nucl.\ Instrum.\
  Meth.\  \textbf{A506} (2003) 250}\relax
\mciteBstWouldAddEndPuncttrue
\mciteSetBstMidEndSepPunct{\mcitedefaultmidpunct}
{\mcitedefaultendpunct}{\mcitedefaultseppunct}\relax
\EndOfBibitem
\bibitem{LHCb-PROC-2011-006}
M.~Clemencic {\em et~al.}, \ifthenelse{\boolean{articletitles}}{\emph{{The
  \lhcb simulation application, Gauss: Design, evolution and experience}},
  }{}\href{https://doi.org/10.1088/1742-6596/331/3/032023}{J.\ Phys.\ Conf.\
  Ser.\  \textbf{331} (2011) 032023}\relax
\mciteBstWouldAddEndPuncttrue
\mciteSetBstMidEndSepPunct{\mcitedefaultmidpunct}
{\mcitedefaultendpunct}{\mcitedefaultseppunct}\relax
\EndOfBibitem
\bibitem{Cowan:2016tnm}
G.~A. Cowan, D.~C. Craik, and M.~D. Needham,
  \ifthenelse{\boolean{articletitles}}{\emph{{RapidSim: an application for the
  fast simulation of heavy-quark hadron decays}},
  }{}\href{https://doi.org/10.1016/j.cpc.2017.01.029}{Comput.\ Phys.\ Commun.\
  \textbf{214} (2017) 239},
  \href{http://arxiv.org/abs/1612.07489}{{\normalfont\ttfamily
  arXiv:1612.07489}}\relax
\mciteBstWouldAddEndPuncttrue
\mciteSetBstMidEndSepPunct{\mcitedefaultmidpunct}
{\mcitedefaultendpunct}{\mcitedefaultseppunct}\relax
\EndOfBibitem
\bibitem{Hocker:2007ht}
H.~Voss, A.~Hoecker, J.~Stelzer, and F.~Tegenfeldt,
  \ifthenelse{\boolean{articletitles}}{\emph{{TMVA - Toolkit for Multivariate
  Data Analysis with ROOT}}, }{}\href{https://doi.org/10.22323/1.050.0040}{PoS
  \textbf{ACAT} (2007) 040}\relax
\mciteBstWouldAddEndPuncttrue
\mciteSetBstMidEndSepPunct{\mcitedefaultmidpunct}
{\mcitedefaultendpunct}{\mcitedefaultseppunct}\relax
\EndOfBibitem
\bibitem{TMVA4}
A.~Hoecker {\em et~al.}, \ifthenelse{\boolean{articletitles}}{\emph{{TMVA 4 ---
  Toolkit for Multivariate Data Analysis with ROOT. Users Guide.}},
  }{}\href{http://arxiv.org/abs/physics/0703039}{{\normalfont\ttfamily
  arXiv:physics/0703039}}\relax
\mciteBstWouldAddEndPuncttrue
\mciteSetBstMidEndSepPunct{\mcitedefaultmidpunct}
{\mcitedefaultendpunct}{\mcitedefaultseppunct}\relax
\EndOfBibitem
\bibitem{Roe_2005}
B.~P. Roe {\em et~al.}, \ifthenelse{\boolean{articletitles}}{\emph{Boosted
  decision trees as an alternative to artificial neural networks for particle
  identification}, }{}\href{https://doi.org/10.1016/j.nima.2004.12.018}{Nucl.\
  Instrum.\ Meth.\  \textbf{A543} (2005) 577},
  \href{http://arxiv.org/abs/physics/0408124}{{\normalfont\ttfamily
  arXiv:physics/0408124}}\relax
\mciteBstWouldAddEndPuncttrue
\mciteSetBstMidEndSepPunct{\mcitedefaultmidpunct}
{\mcitedefaultendpunct}{\mcitedefaultseppunct}\relax
\EndOfBibitem
\bibitem{Skwarnicki:1986xj}
T.~Skwarnicki, {\em {A study of the radiative cascade transitions between the
  Upsilon-prime and Upsilon resonances}}, PhD thesis, Institute of Nuclear
  Physics, Krakow, 1986,
  {\href{http://inspirehep.net/record/230779/}{DESY-F31-86-02}}\relax
\mciteBstWouldAddEndPuncttrue
\mciteSetBstMidEndSepPunct{\mcitedefaultmidpunct}
{\mcitedefaultendpunct}{\mcitedefaultseppunct}\relax
\EndOfBibitem
\bibitem{LHCb-PAPER-2022-010}
LHCb collaboration, R.~Aaij {\em et~al.},
  \ifthenelse{\boolean{articletitles}}{\emph{{Measurement of $\tau_L$ using the
  $\Bs \rightarrow \jpsi \etaz$ decay mode}},
  }{}\href{https://doi.org/10.1140/epjc/s10052-023-11634-4}{Eur.\ Phys.\ J.\
  \textbf{C83} (2023) 629},
  \href{http://arxiv.org/abs/2206.03088}{{\normalfont\ttfamily
  arXiv:2206.03088}}\relax
\mciteBstWouldAddEndPuncttrue
\mciteSetBstMidEndSepPunct{\mcitedefaultmidpunct}
{\mcitedefaultendpunct}{\mcitedefaultseppunct}\relax
\EndOfBibitem
\bibitem{LHCb-PAPER-2013-065}
LHCb collaboration, R.~Aaij {\em et~al.},
  \ifthenelse{\boolean{articletitles}}{\emph{{Measurements of the \Bp, \Bz, \Bs
  meson and \Lb baryon lifetimes}},
  }{}\href{https://doi.org/10.1007/JHEP04(2014)114}{JHEP \textbf{04} (2014)
  114}, \href{http://arxiv.org/abs/1402.2554}{{\normalfont\ttfamily
  arXiv:1402.2554}}\relax
\mciteBstWouldAddEndPuncttrue
\mciteSetBstMidEndSepPunct{\mcitedefaultmidpunct}
{\mcitedefaultendpunct}{\mcitedefaultseppunct}\relax
\EndOfBibitem
\bibitem{LHCb-PAPER-2012-005}
LHCb collaboration, R.~Aaij {\em et~al.},
  \ifthenelse{\boolean{articletitles}}{\emph{{Analysis of the resonant
  components in \mbox{\decay{\Bsb}{\jpsi\pip\pim}}}},
  }{}\href{https://doi.org/10.1103/PhysRevD.86.052006}{Phys.\ Rev.\
  \textbf{D86} (2012) 052006},
  \href{http://arxiv.org/abs/1204.5643}{{\normalfont\ttfamily
  arXiv:1204.5643}}\relax
\mciteBstWouldAddEndPuncttrue
\mciteSetBstMidEndSepPunct{\mcitedefaultmidpunct}
{\mcitedefaultendpunct}{\mcitedefaultseppunct}\relax
\EndOfBibitem
\end{mcitethebibliography}
\end{document}